\documentclass[aps,preprint,prd,eqsecnum,nofootinbib,showpacs,12pt]{revtex4}

\usepackage{epsfig}
\usepackage{bm}

\newcommand{\ens}{{\cal E}}
\newcommand{\boa}{{\bm{\alpha}}}

\newcommand\half{{\textstyle{1\over2}}}
\newcommand\goto{\mathop{\longrightarrow}}
\newcommand\Tr{\mathop{\rm Tr}\nolimits}
\newcommand{\gs}{\Psi_0^{(1)}}
\newcommand\bn{\bm{n}}

\begin{document}

\date{February 9, 2004}

\title{
{\vspace{-1.2em} \parbox{\hsize}{\hbox to \hsize 
{\hss \normalsize\rm IFUP-TH 2004/09, UPRF-2004-02}}} \\
Supersymmetry breaking in two dimensions: \\
the lattice $\bm{N=1}$ Wess-Zumino model}

\author{Matteo Beccaria}%
\email{Matteo.Beccaria@le.infn.it}%
\affiliation{INFN, Sezione di Lecce, and
Dipartimento di Fisica dell'Universit\`a di Lecce,
Via Arnesano, ex Collegio Fiorini, I-73100 Lecce, Italy}

\author{Massimo Campostrini}%
\email{Massimo.Campostrini@df.unipi.it}%
\affiliation{INFN, Sezione di Pisa, and
Dipartimento di Fisica ``Enrico Fermi'' dell'Universit\`a di Pisa,
Via Buonarroti 2, I-56125 Pisa, Italy}

\author{Alessandra Feo}%
\email{feo@fis.unipr.it}%
\affiliation{Dipartimento di Fisica, Universit\`a di Parma 
and INFN Gruppo Collegato di Parma,
Parco Area delle Scienze, 7/A, 43100 Parma, Italy}

\begin{abstract}
We study dynamical supersymmetry breaking by non perturbative lattice
techniques in a class of two-dimensional $N=1$ Wess-Zumino models.  We
work in the Hamiltonian formalism and analyze the phase diagram by
analytical strong-coupling expansions and explicit numerical
simulations with Green Function Monte Carlo methods.
\end{abstract}


\maketitle


\section{Introduction}
\label{Sec:Intro}

Supersymmetry (SUSY) is playing an increasingly relevant and unifying
role in high energy physics. From a purely theoretical point of view,
SUSY is required for consistency and finiteness in superstring theory;
compactification and SUSY breaking mechanisms are then needed in order
to produce a low energy four-dimensional effective action with a
residual $N=1$ SUSY.  This constraint comes from the phenomenological
side where the most popular current extensions of the Standard Model
are actually based on SUSY for at least two reasons.  First,
supersymmetric Grand Unification theories are quite successful in
predicting $SU(3)\times SU(2)\times U(1)$ gauge couplings
unification~\cite{Unification}, a fact that can be considered as the
main phenomenological motivation for SUSY~\cite{DESY}.  Moreover,
supersymmetric models solve in a natural way the hierarchy
problem~\cite{Hierarchy} of matching the electroweak and GUT scales
without being spoiled by huge radiative corrections to Higgs masses.

However, many features of this scenario still need some clarification.
Indeed, $N=1$ SUSY is expected to be exact at the GUT scale around
$10^{16}$ GeV, but must be broken in the TeV region in order to
account for the asymmetric mass textures that are currently observed.
In particular, this will be true if some superpartner with a mass of a
few TeV will be observed in the future LHC or Linear Collider
experiments.  In the model independent analysis, the source of
breaking is usually parametrized by a complete set of soft terms whose
origin remains however rather unexplained.

In several approaches, it is due to some kind of spontaneous breaking
of SUSY in a {\em hidden} sector and communicated to the MSSM
particles producing the soft terms.  As with every dynamical symmetry
breaking, non-perturbative techniques must be exploited and the
lattice regularization and renormalization programme is of course one
of the main lines of research.  Indeed, the simultaneous introduction
of infrared and ultraviolet cutoffs allows for calculations, like
strong-coupling expansions, that are quite complementary to the usual
weak-coupling perturbative analysis.

Beside this, when any known analytical treatment fails, lattice models
can also be studied by direct simulations that can provide, in
principle, accurate numerical measurements.

In this paper, we address the problem of spontaneous supersymmetry
breaking (S${}^3$B) in a simple, but interesting, theoretical
laboratory that is the class of Wess-Zumino (WZ) two-dimensional
models of chiral superfields with no vector multiplets.  Preliminary
results on this subject can be found in~\cite{Previous}.  Related
studies of the two dimensional Wess-Zumino model can be found in
\cite{catterall}.

Despite their simplicity, these systems are nontrivial because in two
dimensions supersymmetry is not strong enough to predict the exact
pattern of breaking, a situation that must be compared with the four
dimensional case where WZ models are expected to break supersymmetry
if and only if they do at tree level.

Unfortunately, as we shall discuss, lattice strong-coupling expansions
provide useful insights, but are unable to reliably predict the
physics of the continuum theory and one must resort to a numerical
analysis.

Since S${}^3$B is closely related to the symmetry properties of the
ground state, it appears to be quite reasonable to adopt some kind of
Hamiltonian formulation.  Moreover, we will see in the following that,
if we wish to preserve a SUSY subalgebra, a conserved Hamiltonian is
crucial.  However, the traditional algorithms for simulation of
lattice field theories are based on the Lagrangian
formulation~\cite{Lagr}.  The main reason is the immediate
probabilistic interpretation of the partition function, at least for
bosonic systems not suffering from a sign or phase problem; this leads
to a host of Monte Carlo algorithms, some of which are extremely
efficient.  Of course, alternatives based on a more direct Hamiltonian
formalism do exist~\cite{KS}, but they are certainly less exploited in
high energy physics where emphasis is on Lagrangian symmetries, in
particular Poincar\'e invariance.

On the other hand, Hamiltonian methods has been used in Supersymmetric
Discretized Light-Cone Quantization (SDLCQ)~\cite{Pinsky} and also are
widely exploited in non relativistic contexts~\cite{IS} where the
properties of the ground state are typically the simplest and first
object of investigation.  Moreover, these techniques interlace the
brute force numerical calculations with analytical or physical
insights about the structure of the ground state wave function, a
feature that is quite welcome in the study of S${}^3$B where we expect
major changes to show up at the phase transition.

Another important feature of our study concerns the fact that
fermions, needed in supersymmetric models, are the major source of
complications in the current approaches to lattice simulations. In the
Lagrangian approach quantum expectation values are computed by summing
over histories of the classical fields, following Feynman's ideas.  In
the case of fermions, these are Grassmann valued {\em classical}
fields that cannot be analyzed by direct numerical methods. The
typical solution amounts to integrate them out and study the resulting
non-local bosonic model~\cite{Montvay}. This can be nontrivial for a
generic model and a recent detailed account of this problem and
whether it can be formulate successfully supersymmetric theories on
the lattice can be found in~\cite{Ale}.

Instead, in the Hamiltonian approach, what is
relevant is the algebra of the fields and their conjugate
momenta. From this point of view, fermions and bosons differ just by
the replacement of commutators by anticommutators and also by the
structure of the state space, finite dimensional for fermions in
finite volume, infinite dimensional in the bosonic case. Apparently, 
in the Hamiltonian approach, there is much more symmetry in the
treatment of fermions and bosons than in the Lagrangian approach. 

Looking at the details of the simulation techniques, however, problems
arise with Hamiltonian fermions due to the well known hard sign
problem~\cite{GenericSignProblem}.  Roughly speaking, fermion
exchanges introduce problematic and unavoidable signs that often break
in a substantial way the probabilistic interpretation of quantum
expectation values required to build a numerical stochastic algorithm.
This deep problem is somewhat milder in $1+1$ dimensions where
specific equivalences between fermionic and bosonic fields can be
established~\cite{Bosonization}.  Also, the topology of fermion
dynamics is the simplest possible and helps in taming the sign
problem. Actually, for several fermion models in $1+1$ dimensions
arising in Solid State theory, like, e.g., Hubbard-like models,
algorithms can be devised with no sign problem and good efficiency as
well as scaling properties~\cite{Linden}.

The detailed plan of the paper is the following: in
Sec.~(\ref{Sec:Model}) we present the model and its lattice
Hamiltonian; in Sec.~(\ref{Sec:SC}) we compute the first nontrivial
order of the strong-coupling expansion of the ground state energy; in
Sec.~(\ref{Sec:RG}) we discuss the Renormalization Group trajectories
along which a continuum limit can be reached.  In
Sec.~(\ref{Sec:Simulation}) we describe a Green Function Monte Carlo
algorithm. Finally, Sec.~(\ref{Sec:Results}) is devoted to present our
numerical results.


\section{The $\bm{N=1}$ Wess-Zumino model in $\bm{1+1}$ dimensions}
\label{Sec:Model}

\subsection{Definition of the model and patterns of SUSY breaking}

Let us consider the most general SUSY algebra in two dimensions. The
generators are split into fermionic and bosonic ones. The algebra with
$N$ left-handed fermionic generators $\{Q_L^A\}_{A=1,\dots, N}$ and
$\bar N$ right-handed fermionic generators $\{Q_R^A\}_{A=1,\dots,\bar N}$
is denoted by $(N, \bar N)$. The bosonic generators are the
components of the two-momentum $(P^0, P^1)$ and a set of central
charges $T^{AB}$. The non trivial part of the algebra is 
\begin{eqnarray}
\{Q_L^A, Q_L^B\} &=& \delta^{AB}(P^0-P^1) , \nonumber\\
\{Q_R^A, Q_R^B\} &=& \delta^{AB}(P^0+P^1) , \nonumber\\
\{Q_L^A, Q_R^B\} &=& T^{AB} . \nonumber 
\end{eqnarray}
In the left-right symmetric case with $(N,\bar N) = (1,1)$, we denote
\begin{equation}
Q_{1,2} \equiv Q_R^1\pm Q_L^1 ,
\end{equation}
and find
\begin{equation}
\label{SUSY}
\{Q_a, Q_b\} = 2(H\ {\bf 1} + P\sigma^1 + T\sigma^3)_{ab},
\end{equation}
where $\sigma^i$ are the Pauli matrices, $(P^0, P^1)\equiv (H, P)$ and
$T \equiv T^{11}$.  The minimal realization of this algebra requires a
single real chiral multiplet with a real scalar component $\varphi$
and a Majorana fermion with components $\psi^{1,2}$.  The explicit
form of the supercharges is
\begin{equation}
Q_{1,2} = \int dx 
\left[p(x)\psi^{1,2}(x)
-\left(\frac{\partial\varphi}{\partial x}
 \pm V(\varphi(x))\right)\psi^{2,1}(x)\right] ,
\end{equation}
where $p(x)$ is the momentum operator conjugate to $\varphi(x)$. The
central charge corresponds to a topological quantum
number~\cite{WittenOlive}
\begin{equation}
T = \int dx \frac{\partial\varphi}{\partial x} V(\varphi).
\end{equation}
As usual with supersymmetric models, the structure of the Hamiltonian
$H$ guarantees that the energy eigenstates have $E\ge 0$ because
\begin{equation}
H = \frac{1}{2}(Q_1^2+Q_2^2) .
\end{equation}
Invariant states annihilated by both $Q_i$ coincide with the zero
energy states and are thus supersymmetric ground states; they must lie
in the topologically trivial sector.

The problem of predicting the pattern of $S^3B$ is not easy. In
principle, the form of $V(\varphi)$ is enough to determine whether
supersymmetry is broken or not. At least at tree level, one easily
proves that supersymmetry is broken if and only if $V$ has no zeros.
In two dimensions, however, these conclusion is generally false due to
radiative corrections. An analytic non-perturbative tool that can help
in the analysis is the Witten index defined as~\cite{Witten82}
\begin{equation}
I = \mbox{Tr}(-1)^F,
\end{equation}
where $F$ is the fermion number. Since supersymmetry is not explicitly
broken, contributions from positive-energy bosonic and fermionic
states cancel and
\begin{equation}
I = n^B_{E=0}-n^F_{E=0} .
\end{equation}
In finite volume, $I$ is invariant under changes in $V(\varphi)$ that
do not modify its asymptotic behaviour. In particular, it can be
computed at weak coupling where each zero of $V(\varphi)$ is
associated to a perturbative zero energy state. Thus, if $V(\varphi)$
has an odd number of zeroes, we find $I\neq 0$ and supersymmetry is
unbroken.  If, on the other hand, $V(\varphi)$ has an even number of
zeroes, the associated perturbative vacua can contribute $I$ with
opposite signs and, when $I=0$, we cannot say anything. In particular,
a nontrivial set of perturbative zero energy states with $I=0$ can
receive instanton corrections due to tunneling lifting them to
positive energies breaking supersymmetry while leaving $I=0$.  In such
cases, the behaviour of the tunneling rate with increasing volume is
crucial in answering the question of breaking.

An interesting example of this complicated scenario is discussed in
Appendix A of Ref.\ \cite{Witten82}. We quickly review the analysis
since it will be important in the interpretation of our results.  When
$V(\varphi) = \lambda(\varphi^2 + a^2)$, the action of the WZ model is
\begin{equation}
S = \int d^2x\left(\frac 1 2 (\partial\varphi)^2
     + \frac{i}{2}\bar\psi\gamma\cdot \partial\psi
     - \frac 1 2 \lambda^2(\varphi^2+a^2)^2
     - \frac 1 2 \lambda\varphi\bar\psi\psi\right).
\end{equation}
For large positive $a^2$ the index is zero because there are no
zero-energy states. Due to a special conjugation symmetry valid for
this model in finite volume, the pattern of breaking is invariant
under $a^2\to - a^2$.  This means that for negative $a^2$, the two
zeroes of $V$ are bosonic and fermionic and (finite volume) tunneling
lifts their energy to a positive value. However, in infinite volume
and large negative $a^2$, the narrow minimum of the potential is
protected from radiative corrections and generates an expectation
value $\langle\varphi\rangle\neq 0$ signaling the SSB of the $Z_2$
symmmetry $\varphi\to-\varphi$, $\psi\to\gamma_5\psi$.  The fermion
becomes massive and supersymmetry must be unbroken due to the absence
of a massless Goldstino.

The above discussion illustrates that an alternative non-perturbative
analysis of the models with $I=0$ is certainly welcome. In the
following, we shall put the model on a space-time lattice in order to
exploit explicit numerical simulations as well as analytical
strong-coupling expansions.

\subsection{Lattice Version of the Model}

On the lattice it is impossible to maintain the full SUSY algebra and
it is important to understand what can be said by looking at
subalgebras.  If we consider one supercharge only, for instance $Q_1$,
and find a state with $Q_1|s\rangle = 0$, we cannot say that it is a
zero energy state unless we know that it is in the $T=0$ sector.  On
the other hand, if no states with $Q_1|s\rangle = 0$ are found in any
topological sector, then supersymmetry is certainly broken.

Thus, even if we forget $Q_2$, we can choose as a clear-cut signal of
supersymmetry dynamically breaking the lowest eigenvalue of the
operator $Q_1^2$: if it is positive, we have breaking.

The SUSY algebra (\ref{SUSY}) involves explicitely the generators of
space and time infinitesimal translations and is spoiled on the
lattice.  In the Lagrangian approach, both space and time are discrete
and SUSY is completely broken. In the Hamiltonian formulation, time
remains continuous and the $D=2$ algebra is reduced to $D=1$ and not
totally lost. The full two-dimensional algebra as well as Lorentz
invariance are expected to be recovered in the continuum limit.

A lattice version of the above model has been previously studied in
Refs.\ \cite{ERS,Ranft-Schiller}.  A similar approach to Wess-Zumino
models with $N=2$ supersymmetry is discussed in Ref.\ 
\cite{Elitzur-Schwimmer}, and numerical investigations are reported in
Ref.\ \cite{Schiller-Ranft}.  On each site of a spatial lattice with
$L$ sites, we define a real scalar field $\varphi_n$ together with
its conjugate momentum $p_n$ such that $[p_n, \varphi_m] =
-i\delta_{n,m}$.  The associated fermion is a Majorana fermion
$\psi_{a, n}$ with $a=1, 2$ and $\{\psi_{a, n}, \psi_{b, m}\} =
\delta_{a,b}\delta_{n,m}$ , $\psi_{a,n}^\dagger = \psi_{a,n}$. The
fermionic charge
\[
Q = \sum_{n=1}^L\left [
p_n\psi_{1,n}-\left(\frac{\varphi_{n+1}-\varphi_{n-1}}{2}
+V(\varphi_n)\right)\psi_{2,n}\right],
\]
with arbitrary real function $V(\varphi)$, (called {\em prepotential}
in the following) can be used to define a semi-positive definite
lattice Hamiltonian
\begin{equation}
H = Q^2.
\end{equation}
This Hamiltonian includes the central charge contribution in the form
of a term
\begin{equation}
\sum_n V(\varphi_n)\frac{\varphi_{n+1}-\varphi_{n-1}}{2} ,
\end{equation}
that is precisely a discretization of $T$.  Eigenstates of $H$ are
divided into invariant $Q$-singlets with zero energy and $Q$-doublets
with degenerate positive energy.  $H$ describes an interacting model,
symmetric with respect to $Q$ and this symmetry is respected by the
spectrum if and only if the ground state energy is positive.  We
stress again that $Q$ symmetry breaking implies breaking of the full 2
dimensional supersymmetry, whereas $Q$ symmetry does not imply (in a
generic topological sector) 2D SUSY.

We remind that, on the lattice, spontaneous supersymmetry breaking can
occur even for finite lattice size $L$, because tunneling among
degenerate vacua connected by $Q$ is forbidden by fermion number
conservation.

To write $H$ in a more familiar form, we follow Ref.\ 
\cite{Ranft-Schiller} and replace the two Majorana fermion operators
with a single Dirac operator $\chi$ satisfying canonical
anticommutation rules, i.e.,
$\{\chi_n, \chi_m\} = 0$, $\{\chi_n,\chi_m^\dagger\} = \delta_{n,m}$:
\begin{equation}
\psi_{1,n} = \frac{(-1)^n-i}{2i^n}(\chi_n^\dagger+i\chi_n) , \qquad
\psi_{2,n} = \frac{(-1)^n+i}{2i^n}(\chi_n^\dagger-i\chi_n) .
\label{eq:chi}
\end{equation}
The Hamiltonian
takes then the form 
\begin{eqnarray}
H &=& H_B(p, \varphi) + H_F(\chi, \chi^\dagger, \varphi) 
\nonumber  \\ 
 &=& \sum_n\left\{ \frac 1 2 p_n^2 + \frac 1
2\left(\frac{\varphi_{n+1}-\varphi_{n-1}}{2} + V(\varphi_n)\right)^2
\right. \nonumber \\
&& \left .  -\frac 1 2 (\chi^\dagger_n\chi_{n+1} + h.c.) +
(-1)^n V'(\varphi_n) \left(\chi^\dagger_n\chi_n-\frac 1 2 \right)
\right\}
\end{eqnarray}
and describes canonical bosonic and fermionic fields with standard
kinetic energies and a Yukawa coupling.

This Hamiltonian conserves the total fermion number
\begin{equation}
N_f = \sum_n \chi^\dagger_n\chi_n ,
\end{equation}
and can be examined in each sector with definite $N_f$ separately. For
reasons that will be understood later, we shall also consider open
boundary conditions and restrict the lattice size $L$ to be even.
These are constraints that do not affect the physics of the model in
the continuum, but will be very welcome by the algorithm we are going
to apply.

The simplest way to analyze the pattern of supersymmetry breaking for
a given $V(\varphi)$ is to compute the ground state energy $E_0$. As
we mentioned, on the lattice, we can perform such a computation in a
non-perturbative way by strong coupling or numerical simulations.
However, before discussing these items, we want to stress some
identities that can be used together with $E_0$ to get information on
the symmetry of the ground state.

\subsection{Lattice Ward identities}

If the vacuum $|0\rangle$ is supersymmetric, $Q|0\rangle = 0$ and for
each operator $X$ we have
\begin{equation}
\langle 0 | \{Q, X\} | 0\rangle = 0 .
\end{equation}
In particular, taking
\begin{equation}
X = \sum_n F(\varphi_n) \psi_{2, n},
\end{equation}
we obtain 
\begin{equation}
\langle 0 | \sum_n\left\{ 
F(\varphi_n)\left[\frac {\varphi_{n+1}-\varphi_{n-1}}{2} + V(\varphi_n)
\right] 
+ F'(\varphi_n) (-1)^n (\chi_n^\dagger\chi_n -1/2)\right\}
| 0\rangle = 0 .
\end{equation}

A basis of Ward Identities is thus obtained by considering $F(\varphi)
= \varphi^n$. For instance, on an even lattice with open boundary
conditions we find for $n=1$ the relation
\begin{equation}
\langle 0 | \sum_n \left\{ 
\varphi_n V(\varphi_n) 
+ (-1)^n \chi_n^\dagger\chi_n \right\}
| 0\rangle = 0 .
\end{equation}
The case $F(\varphi)=\mbox{constant}$ is also interesting. It leads to 
\begin{equation}
\langle 0 | \sum_n V(\varphi_n) | 0 \rangle = 0 .
\end{equation}


\section{Strong coupling analysis of SUSY breaking}
\label{Sec:SC}

Let us start from the supersymmetry charge
\[
Q = \sum_l \left[p_l\psi^1_l - V(\varphi_l)\,\psi^2_l
- {\varphi_{l+1} - \varphi_{l-1} \over 2}\,\psi^2_l\right],
\]

Following Ref.\ \cite{Elitzur-Schwimmer}, we define the
strong-coupling limit by
\[
V(\varphi) \goto_{\lambda\to\infty} \lambda V^{(0)}(\lambda \varphi),
\]
perform the canonical transformation
\[
\varphi^{(0)} = \lambda\varphi, \quad p^{(0)} = {1\over\lambda}\,p,
\]
and rescale the energies by $\lambda^2$; dropping the index $^{(0)}$
from $\varphi$ and $p$, the result is
\begin{eqnarray*}
&&Q =  \sum_l \left[p_l\psi^1_l - V(\varphi_l)\,\psi^2_l
- {(\varphi_{l+1} - \varphi_{l-1}) \psi^2_l \over 2\lambda^2}\right]
\equiv Q^{(0)} + {Q^{(2)}\over\lambda^2},
 \\
&&H = Q^2 =
  {1\over2}\sum_l \Biggl[p_l^2 + V^2(\varphi_l) 
   + 2 i V'(\varphi_l)\,\psi_l^1\psi_l^2 \\
&&\qquad +\, {(\varphi_{l+1} - \varphi_{l-1}) V(\varphi_l)
      + i\psi_{l+1}^1\psi_l^2 + i \psi_{l+1}^2 \psi_l^1 \over \lambda^2}
   + {(\varphi_{l+1}-\varphi_{l-1})^2\over 4\lambda^4}\Biggr]
\equiv H^{(0)} + {H^{(2)}\over\lambda^2} + {H^{(4)}\over\lambda^4}.
\end{eqnarray*}

Introducing the Dirac fields $\chi_l$, $\chi^\dagger_l$
\cite{Ranft-Schiller}, cf.\ Eq.\ (\ref{eq:chi}), we obtain 
\begin{eqnarray*}
H^{(0)} &=& \sum_l\left[
\frac 1 2 p_l^2 + \frac 1 2 V^2(\varphi_l)+(-1)^l V'(\varphi_l)
(\chi_l^\dagger \chi_l-1/2)
\right] = \\
&=& \sum_l\left[
\frac 1 2 p_l^2 + \frac 1 2 V^2(\varphi_l)+\frac 1 2 (-1)^{l+n_l} 
V'(\varphi_l) \right] \\
H^{(2)} &=& \frac{1}{2}\sum_l V(\varphi_l)(\varphi_{l+1}-\varphi_{l-1})
-\frac{1}{2}\sum_l(\chi^\dagger_l\chi_{l+1} + \mbox{h.c.})\\
H^{(4)} &=& \frac 1 8 \sum_l(\varphi_{l+1}-\varphi_{l-1})^2
\end{eqnarray*}
where we denote by $n_l=0,1$ the eigenvalue of $\chi^\dagger_l \chi_l$.

\subsection{Leading order}

To leading order in $1/\lambda$, the Hamiltonian is factorized in a
supersymmetric quantum mechanics for each site; adopting an explicit
representation, we can write the one-site Hamiltonian as
\begin{equation}
H = \frac 1 2 \left[- {d^2\over d x^2} + V^2(x) + \sigma_3 V'(x) \right]
\label{eq:H0-repr}
\end{equation}
(in the occupation number representation the basis chosen now is
$(n{=}0,n{=}1)$ for odd sites and $(n{=}1,n{=}0)$ for even sites).
This Hamiltonian has a $N=2$ supersymmetry \cite{Witten81}:
\begin{equation}
\{Q_i,Q_j\} = \delta_{ij} H, \qquad
Q_1 = \half \left[\sigma_1 p + \sigma_2 V(x)\right], \qquad
Q_2 = \half \left[\sigma_2 p - \sigma_1 V(x)\right].
\label{eq:algebra}
\end{equation}
The conditions for a supersymmetric ground state $Q_i\psi_0=0$ reduce
to
\begin{equation}
\psi'_0(x) = \sigma_3 V(x) \, \psi_0(x).
\label{eq:psi0}
\end{equation}
For polynomial $V$, supersymmetry is unbroken if and only if it is
possible to find a normalizable solution to Eq.\ (\ref{eq:psi0}),
which happens if the degree $q$ of $V$ is odd \cite{Witten81}.

We can write the time-independent Schr\"odinger equation as
\[
\psi'' + 
\left(2E - V^2(x) \mp V'(x) \right) \psi = 0;
\]
denoting the eigenfunctions for the two signs by
$\psi_m^\pm$ and their energies by $\varepsilon_m^\pm$, we have
\begin{equation}
{\psi_m^\pm}'' + \left(2 \varepsilon_m^\pm - 
        V^2(x) \mp V'(x) \right) \psi_m^\pm = 0.
\label{eq:psi-pm}
\end{equation}
Supersymmetry implies that, for $E\ne0$, states are paired in
boson-fermion doublets. 

We remark that this conclusion is in strong disagreement with the
continuum (or weak-coupling lattice) analysis where the relevant
feature of $V$ is the existence of zeros.

\begin{figure}[tbp]
\begin{center}
\leavevmode
\epsfig{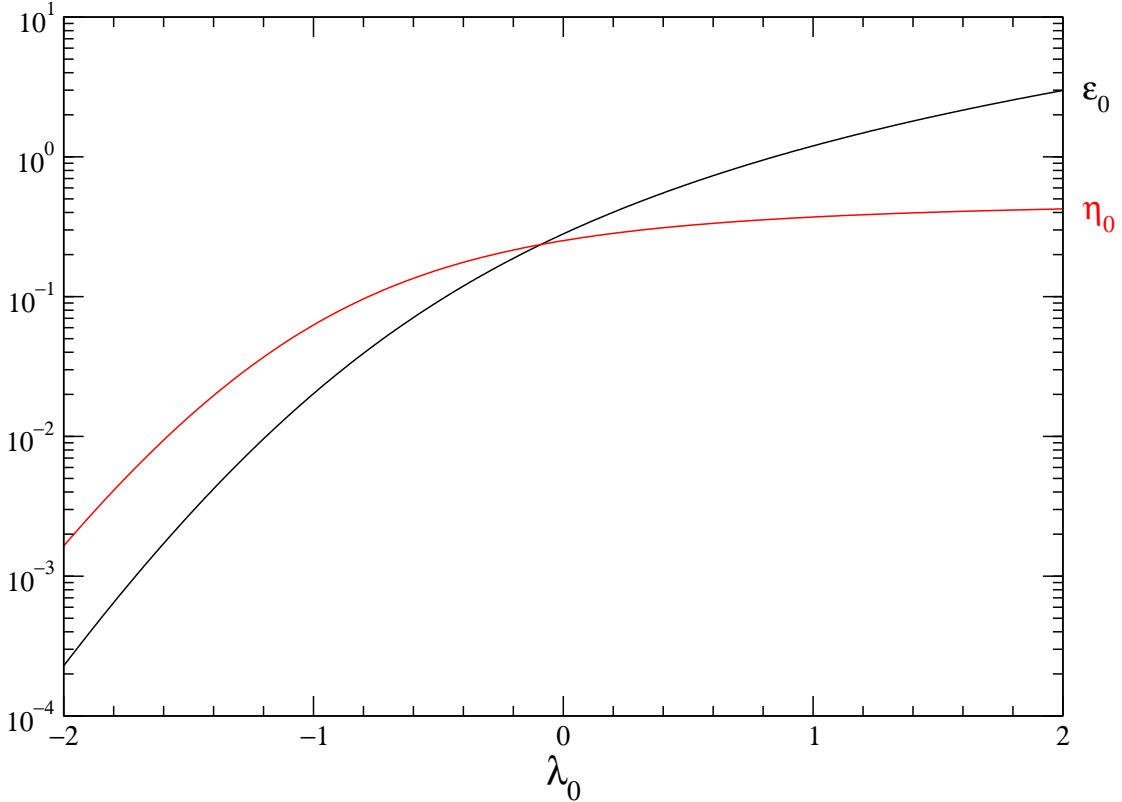}
\vskip-7mm
\end{center}
\caption{The ground-state energy $\varepsilon_0=\varepsilon_0^\pm$ and
the overlap $\eta_0$ as functions of $\lambda_0$ for the quadratic
prepotential $V = \varphi^2 + \lambda_0$.
}
\label{fig:epsilon+eta}
\end{figure}

For $q>1$, $\psi_m^\pm(x)$ and $\varepsilon_m^\pm$ cannot be computed
exactly (excluding the cases $\varepsilon_0^\pm=0$); it is however
easy to compute then numerically to high accuracy, using, e.g., the
Numerov method \cite{Johnson}.  An example is shown in Fig.\ 
\ref{fig:epsilon+eta}.

In the following analysis, we shall have to tell between $V$ with odd
or even leading power of $\varphi$.

\subsubsection{Odd $q$}

For odd $q$, we either have $\varepsilon_0^-=0$ or
$\varepsilon_0^+=0$; let us assume $\varepsilon_0^-=0$: 
$\psi_0^-$ is the supersymmetric ground state satisfying Eq.\ 
(\ref{eq:psi0}); all the other states appear in pairs: $\varepsilon_{m+1}^- =
\varepsilon_m^+$.  Notice that the ground state is bosonic for even sites,
fermionic for odd sites (the opposite holds if $\varepsilon_0^+=0$).  

A strong argument against supersymmetry breaking is given by the
Witten index $I_W \equiv \Tr(-1)^{N_f}$ \cite{Witten82}; in the
strong-coupling limit, we clearly have $I_W=\pm1$; since $I_W\ne0$
implies unbroken SUSY, and $I_W$ is invariant under regular
perturbations (cf.\ Sect.\ \ref{regular-pert}), supersymmetry can
never be broken for odd $q$, not even in the $L\to\infty$ limit.

A simple check of this result can be given explicitely when $V$ is
linear and is discussed in details in
App.~(\ref{App:LinearPotential}).

\subsubsection{Even $q$}

For even $q$, we have $\varepsilon_m^+ = \varepsilon_m^-$; for $m=0$,
this corresponds to a degenerate ground state with broken
supersymmetry ($\varepsilon_0^\pm = \varepsilon_0 > 0$).  The phases
of the normalized states $|\psi_n^\pm\rangle$ can be chosen in order
to satisfy
\begin{eqnarray}
{\textstyle\sqrt{2\varepsilon_n}}|\psi_n^-\rangle 
    &=& \left[-ip+V(\varphi)\right]|\psi_n^+\rangle,
\label{eq:Qpsip}  \\
{\textstyle\sqrt{2\varepsilon_n}}|\psi_n^+\rangle 
    &=&\left[ip+V(\varphi)\right]|\psi_n^-\rangle;
\label{eq:Qpsim}
\end{eqnarray}
Introducing the notation
\begin{eqnarray*}
&\langle {\cal O}\rangle_\pm = 
\langle \psi_0^\pm | {\cal O} |\psi_0^\pm \rangle , \\
&\langle \psi_0^+ | \psi_0^- \rangle = \eta_0,
\end{eqnarray*}
we can prove the important relations
\begin{eqnarray}
\langle V \rangle_\pm &=& {\textstyle\sqrt{2\varepsilon_0}} \, \eta_0,
\label{eq:Vpm} \\
\langle \varphi\rangle_- - \langle \varphi \rangle_+ 
&=& \    frac{1}{\sqrt{2\varepsilon_0}}
\,\eta_0. \label{eq:xmp}
\end{eqnarray}
$\eta_0$ can be computed numerically from $\psi_0^\pm(\varphi)$, cf.\ Fig.\ 
\ref{fig:epsilon+eta}.
The proof of Eq.\ (\ref{eq:Vpm}) is very simple: just
take the scalar product of Eq.\ (\ref{eq:Qpsip})
with $\langle\psi_0^+|$ and of Eq.\ (\ref{eq:Qpsip}) with
$\langle\psi_0^-|$, and observe that $\langle p\rangle_\pm=0$.
The proof of Eq.\ (\ref{eq:xmp}) is also immediate:
\begin{eqnarray*}
&&{\textstyle\sqrt{2\varepsilon_0}}\,\langle\psi_0^+|\varphi|\psi_0^+\rangle =
  \langle\psi_0^+|\varphi(ip+V)|\psi_0^-\rangle = 
  \langle\psi_0^+|(ip+V)\varphi|\psi_0^-\rangle +
  \langle\psi_0^+|i[\varphi,p]|\psi_0^-\rangle = \nonumber \\
&&{\textstyle\sqrt{2\varepsilon_0}}\,\langle\psi_0^-|\varphi|\psi_0^-\rangle -
  \langle\psi_0^+|\psi_0^-\rangle,
\end{eqnarray*}

Several simplifications can be exploited when $V(\varphi)$ is even.
For an asymptotically positive polynomial $V(\varphi)$ with degree
$q \ge 2$ it is easy to show that~\footnote{ In fact, from their
definition, one can see that $\psi_n^\pm(\varphi)$ have the same sign
when $\varphi\to +\infty$.  Since they are related by a parity
transformation, their relative phase is fixed by the number of nodes.
}
\[
|\psi_n^-\rangle = (-1)^n I |\psi_n^+\rangle
\]
where $I$ is the hermitian parity inversion operator
\[
\langle \varphi | I | \psi\rangle = \langle -\varphi | \psi\rangle
\]
Then, the eigenstates can be characterized by the single equation 
\[
\sqrt{2\varepsilon_n} |\psi_n\rangle = (-1)^n (ip+V) I |\psi_n\rangle
\]
where 
\[
|\psi_n\rangle\equiv |\psi_n^+\rangle
\]
It is easy in this case to obtain generalized relations like the
previous ones.  Let us consider the equation
\[
\sqrt{2\varepsilon_n}\langle\psi_m |I|\psi_n\rangle 
= (-1)^n\langle \psi_m |I(ip+V)I|\psi_n\rangle
= (-1)^n \langle \psi_m |(-ip + V)|\psi_n \rangle
\]
Taking the hermitian conjugate and exchanging $m$ and $n$ we find the
two equations
\begin{eqnarray}
\sqrt{2\varepsilon_n}\langle\psi_n |I|\psi_m\rangle &=& 
(-1)^n \langle \psi_n |(ip + V)|\psi_m \rangle \\
\sqrt{2\varepsilon_m}\langle\psi_n |I|\psi_m\rangle &=& 
(-1)^m \langle \psi_n |(-ip + V)|\psi_m \rangle
\end{eqnarray}
therefore
\[
\langle\psi_n | V(\varphi) | \psi_m\rangle 
= \frac{1}{\sqrt{2}}(\sqrt{\varepsilon_m}(-1)^m+\sqrt{\varepsilon_n}(-1)^n)
\langle\psi_n
|I|\psi_m\rangle
\]
or (exploiting parity)
\begin{equation}
\label{gen1}
\langle\psi_n^\pm | V(\varphi) | \psi_m^\pm\rangle = 
\frac{1}{\sqrt{2}}(\sqrt{\varepsilon_n}+\sqrt{\varepsilon_m}(-1)^{n+m})
\langle
\psi_n^\mp | \psi_m^\pm\rangle
\end{equation}
In a similar way we can compute 
\[
\sqrt{2\varepsilon_n}\langle\psi_m |\varphi |\psi_n\rangle 
= (-1)^n\langle \psi_m |\varphi (ip+V)I|\psi_n\rangle
\]
Taking the hermitian conjugate and exchanging $m$ and $n$ we find the
two equations
\begin{eqnarray}
\sqrt{2\varepsilon_n}\langle\psi_n |\varphi |\psi_m\rangle &=& 
(-1)^n \langle \psi_n |(-ip - V)\varphi I |\psi_m \rangle \\
\sqrt{2\varepsilon_m}\langle\psi_n |\varphi |\psi_m\rangle &=&
(-1)^m \langle \psi_n |\varphi (ip + V) I |\psi_m 
\rangle
\end{eqnarray}
summing the two equations
\[
\sqrt{2}(\sqrt{\varepsilon_n}(-1)^n +
\sqrt{\varepsilon_m}(-1)^m)\langle\psi_n|\varphi|\psi_m\rangle =
\langle\psi_n i[\varphi, p]I|\psi_m\rangle = 
-\langle\psi_n|I|\psi_m\rangle
\]
and (exploiting parity)
\begin{equation}
\label{gen2}
\langle\psi_n^\pm|\varphi|\psi_m^\pm\rangle = 
\mp\frac{1}{\sqrt{2}}\frac{1}{\sqrt{\varepsilon_n}+(-1)^{n+m}
\sqrt{\varepsilon_m}}
\langle \psi_n^\mp|\psi_m^\pm\rangle
\end{equation}
Of course, for $n=m=0$, Eqs.~(\ref{gen1}), (\ref{gen2}) agree with the
previous general results.

\subsubsection{On the convergence of the perturbative expansion}
\label{regular-pert}

The Rayleigh-Schr\"odinger perturbation theory of a Hamiltonian of
the form $H = H_0 + \beta H_1$ is regular (i.e., it has a finite
radius of convergence in $\beta$) if \cite{Reed-Simon}
\begin{equation}
\Vert H_1\Psi \Vert \le a \Vert H_0\Psi \Vert + b \Vert\Psi\Vert
\label{eq:conv}
\end{equation}
uniformly for all state vectors $\Psi$; in our case, Eq.\ 
(\ref{eq:conv}) clearly holds, for both $H^{(2)}$ and $H^{(4)}$,
except for the trivial cases $q\le1$.

\subsection{First order}

At first order (subleading) in the strong-coupling expansion we
consider again the two cases of even or odd $q$.

\subsubsection{Odd $q$}

In the case of unbroken supersymmetry (odd $q$), the subleading
correction to the ground-state energy in the strong-coupling expansion
is zero: the fermionic contribution $i\psi_{l+1}^1\psi_l^2 +
i\psi_{l+1}^2\psi_l^1$ has clearly zero diagonal matrix elements, and
the bosonic contribution 
$\varphi_{l+1}V(\varphi_l) - \varphi_l V(\varphi_{l+1})$ is zero
because it factorizes into $\langle\varphi\rangle\langle V\rangle -
\langle\varphi\rangle\langle V\rangle$; strictly speaking, this is
true for periodic and free boundary conditions, but it could be false,
e.g., for fixed boundary conditions.

\subsubsection{Even $q$}

Due to the structure of the Hamiltonian, it is convenient to describe
states in the mixed form
\begin{equation}
\sum_{n_1, \dots, n_L} \psi_{n_1, \dots, n_L}(\varphi_1, \dots, \varphi_L)
|n_1, \dots, n_L\rangle 
\end{equation}
where $\psi_{n_1, \dots, n_L}(\varphi_1, \dots, \varphi_L)$ is a wave
function depending on the bosonic degrees of freedom and $ |n_1,
\dots, n_L\rangle $ is the fermionic component of the state defined in
terms of the state annihilated by all $\chi$
\begin{equation}
\chi_i |0\rangle = 0,
\end{equation}
according to the canonical ordering of the Fermi operators:
\begin{equation}
|n_1, \dots, n_L\rangle = 
(\chi_1^\dagger)^{n_1}\cdots (\chi_L^\dagger)^{n_L} |0\rangle .
\end{equation}
Of course $n_i=0,1$ and $|n_1, \dots, n_L\rangle$ describes a state
with $n_i$ fermions at site $i$.  In the case of broken supersymmetry
(even $q$), the subspace $\cal B$ of lowest leading-order energy is
spanned by the states
\begin{equation}
|\bn\rangle = \psi_0^{\sigma_1}(\varphi_1)\cdots\psi_0^{\sigma_{L}}
(\varphi_{L})\, |n_1, \dots, n_{L}\rangle,
\end{equation}
where $\sigma_l = (-1)^{l+n_l}$ and $\psi_0^{\pm 1}\equiv \psi_0^\pm$.
We have adopted open boundary conditions, corresponding in our
notations to setting $\varphi_0 = \varphi_{L+1} = 0$ and $\psi^a_0 =
\psi^a_{L+1} = 0$ (and therefore $\chi_0 = \chi_{L+1} = 0$).

Since the number of fermions $\sum_l n_l$ is conserved, we can 
impose an additional constraint on $\cal B$ and define 
\[
{\cal B}_N = \{|\bn\rangle,\ \sum_l n_l = N\},
\qquad {\cal B} = {\cal B}_0 \oplus \dots \oplus {\cal B}_L.
\]
We will now prove that, for even $L$, the ground state is doubly
degenerate and lies in the sectors with $N=L/2, L/2-1$.

At first order, we have to diagonalize the operator $H^{(2)}$ over 
${\cal B}_N$. Let us split
\begin{eqnarray}
H^{(2)} &=& H^{(2)}_B + H^{(2)}_F \\
H^{(2)}_B &=& 
\frac{1}{2}\sum_{l=1}^L V(\varphi_l)(\varphi_{l+1}-\varphi_{l-1}) \\
H^{(2)}_F &=& 
-\frac{1}{2}\sum_{l=1}^L(\chi^\dagger_l\chi_{l+1} + \chi^\dagger_{l+1}\chi_l)
\end{eqnarray}
Since
\begin{equation}
\langle \bn' | H_B^{(2)} | \bn \rangle = 
\frac{1}{2}{\textstyle\sqrt{2\varepsilon_0}}\,\eta_0\,\delta_{\bn, \bn'}\ 
\sum_l\left(\langle \varphi_{l+1}\rangle - \langle \varphi_{l-1}\rangle\right)
\end{equation}
we have 
\begin{equation}
\langle \bn' | H_B^{(2)} | \bn \rangle =
-\frac 1 4 \eta_0^2 \delta_{\bn, \bn'}\
\left[(-1)^{n_1}+ (-1)^{n_L}\right]
\end{equation}
where we have exploited
\begin{equation}
\langle \varphi_l\rangle = -(-1)^{l+n_l} \frac{\eta_0}{\sqrt{2\varepsilon_0}},
\end{equation}
that holds for even $V$.
Since $n=0,1$ we can use $(-1)^n = 1-2n$ and write
\begin{equation}
\langle \bn' | H_B^{(2)} | \bn \rangle =
\frac 1 2 \eta_0^2 \delta_{\bn, \bn'}\
(-1+n_1+n_L)
\end{equation}
The matrix elements of $H_F^{(2)}$ are 
\begin{equation}
\langle \bn' | H_F^{(2)} | \bn \rangle = 
-\frac{1}{2} \eta_0^2\, h_{\bn, \bn'}
\end{equation}
where $h_{\bn, \bn'}=1$ if $\bn$ and $\bn'$ are
connected by $H_F^{(2)}$ (i.e. a hopping of one fermion)
and $0$ otherwise.

Thus, we can hide the bosonic wave functions and write an effective 
perturbation acting on purely fermionic states as
\begin{equation}
H^{(2)}_{\rm eff} = \frac{\eta_0^2}{2}
\left(\sum_{i,j=1}^L \chi^\dagger_i M_{ij} \chi_j - \bm{1}\right)
\end{equation}
where $\bm{1}$ is the identity operator and
\begin{equation}
M_{ij} = \left\{\begin{array}{ll}
-1 & |i-j|=1 \\
1 & i=j=1 \ \mbox{or}\ L \\
0 & \mbox{otherwise}
\end{array}\right.
\end{equation}
Since $H_{\rm eff}^{(2)}$ is quadratic, it is convenient to change
operator basis.  Let $v^{(p)}_i$ be the p-th eigenvector of $M$ with
eigenvalue $\lambda^{(p)}$:
\begin{eqnarray}
v^{(p)}_n &=& \sqrt\frac{2-\delta_{p,L}}{L}
    \sin\left[\frac{p\pi}{L}\left(n-\frac 1 2\right)\right] \\
\lambda^{(p)} &=& -2\cos\frac{p\pi}{L}
\end{eqnarray}
They define a (real) unitary matrix
\[
\sum_{p=1}^L v^{(p)}_i v^{(p)}_j = \delta_{ij},\qquad
\sum_{i=1}^L v^{(p)}_i v^{(q)}_i = \delta_{pq}
\]
We can replace the operators $\chi_i$ by the operators $a_i$ defined by
\[
\chi_i = \sum_{p=1}^L v^{(p)}_i a_p ,\qquad
a_p    = \sum_{i=1}^L v^{(p)}_i \chi_i
\]
with 
\[
\{a_p, a^\dagger_q\} = \delta_{pq}
\]
The new form of $H_{\rm eff}^{(2)}$ is 
\begin{equation}
H^{(2)}_{\rm eff} = \frac{\eta_0^2}{2}\left(\sum_{p=1}^L \lambda^{(p)} 
a^\dagger_p a_p - \bm{1}\right)
\end{equation}
The eigenvalues $\lambda^{(1)},\dots ,\lambda^{(L/2-1)}$ are negative
and $\lambda^{(L/2)}=0$; there are thus two degenerate ground states
with $L/2-1$ and $L/2$ fermions.  This is required by supersymmetry:
since $H^{(2)}$ restricted to ${\cal B}$ commutes with $Q^{(0)}$, all
the states must be paired in doublets with $N$ differing by 1.  The
ground state with $L/2$ fermions is
\begin{equation}
|\Psi_0^{(1)} \rangle = \psi_0^{\sigma_1}(\varphi_1)\cdots\psi_0^{\sigma_{L}}
(\varphi_{L})\,  a_1^\dagger\ \cdots a_{L/2}^\dagger |0> 
\end{equation}

To conclude, the shift of the ground state energy due to the
perturbation is
\begin{equation}
E_1 = \frac{\eta_0^2}{2}\left(-1-2\sum_{n=1}^{L/2}\cos\frac{\pi n}{L}\right)
= -\frac{\eta_0^2}{2}\cot\frac{\pi}{2L}
\end{equation}
In the $L\to\infty$ limit we have 
\begin{equation}
\frac{E_1}{L} = -\frac{\eta_0^2}{\pi} + {\cal O}(1/L)
\end{equation}
In summary, the first order perturbation in the strong-coupling
expansion removes the large degeneracy of the ground state and
determines a  doublet of eigenstates with $L/2-1$ and $L/2$ fermions
with minimum energy
\begin{equation}
\frac{E}{L} = \varepsilon_0 -\frac{1}{\lambda^2}\frac{\eta_0^2}{\pi} + 
{\cal O}\left(\frac{1}{\lambda^2 L}, \frac{1}{\lambda^4}\right)
\end{equation}
A similar calculation at first order for $\langle \varphi_k \rangle$
and $\langle \varphi_k\varphi_l \rangle_c$ is discussed in
App.~(\ref{App:First}).  The second-order correction to the ground
state energy can also be computed with a reasonable effort and the
result is described in App.~(\ref{App:Second}).  However, we remark
that the results drawn from the first order corrections are not
qualitatively changed.

\subsection{Discussion}

From the analysis of the strong-coupling expansion of the model we can
draw the following conclusion. For polynomial $V(\varphi)$, the
relevant parameter is just its degree $q$.

For odd $q$, the strong coupling analysis and the tree-level results
agree and supersymmetry is expected to be unbroken.  This conclusion
gains further support from the nonvanishing value of the Witten index
at strong coupling.

For even $q$, in strong coupling, the ground state (at least in the
sector with half filling) has a positive energy density also for $L\to
\infty$ and supersymmetry appears to be broken. Of course, this can be
in disagreement with weak coupling. A specific case that we shall
analyze numerically in great detail is
\begin{equation}
V(\varphi) = \lambda_2 \varphi^2 + \lambda_0.
\end{equation}
For $\lambda_0<0$, weak coupling predicts unbroken SUSY, whereas the
strong coupling prediction gives broken SUSY for all $\lambda_0$.  For
this specific model, as we already discussed, the strong coupling
analysis agrees with Ref.\ \cite{Witten82} in the sense that it
reproduces the continuum physics in finite volume.

\begin{figure}[tbp]
\begin{center}
\leavevmode
\epsfig{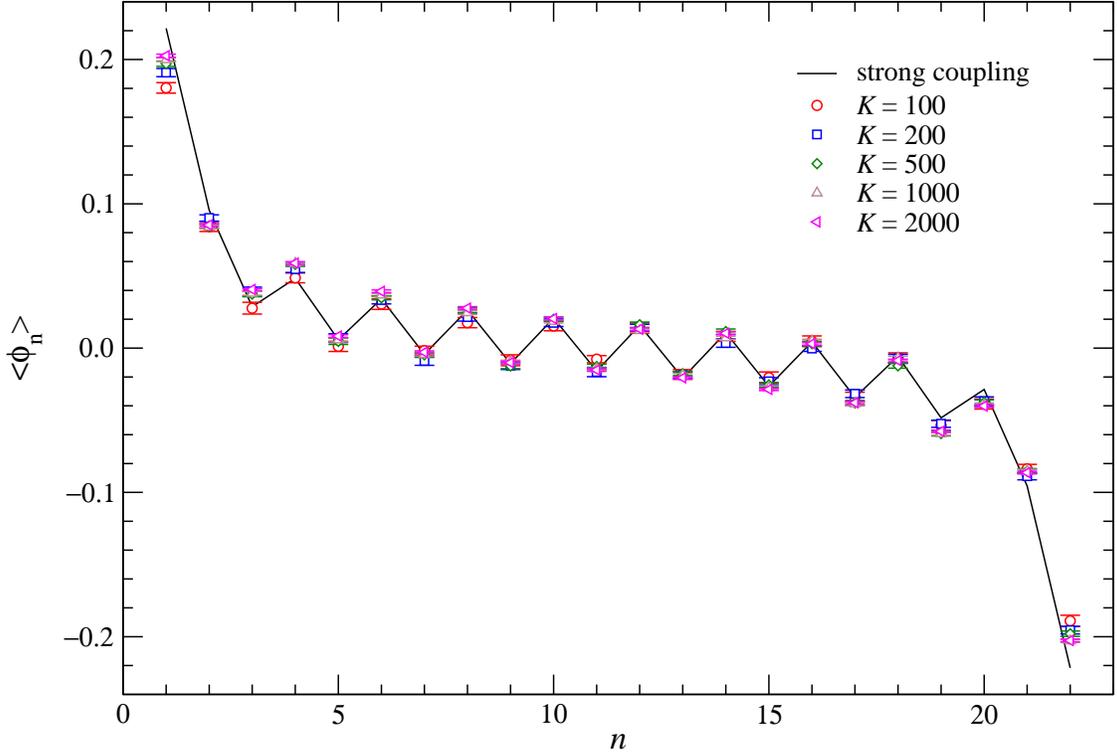}
\vskip-7mm
\end{center}
\caption{Comparison between strong coupling and MC simulation for the
expectation value $\langle\varphi_n\rangle$ in the quadratic model
with $V(\varphi)= 2\varphi^2$ on a lattice with $L=22$ spatial sites.}
\label{fig:SCvsMCphi}
\end{figure}

\begin{figure}[tbp]
\begin{center}
\leavevmode
\epsfig{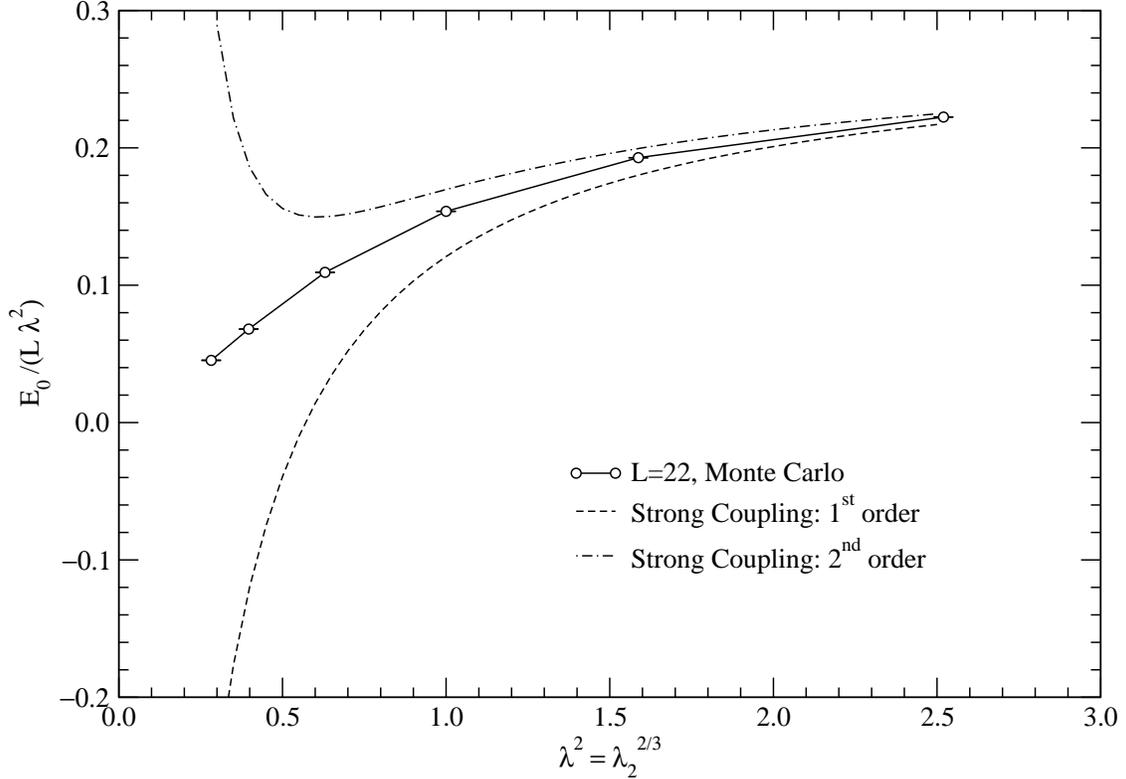}
\vskip-7mm
\end{center}
\caption{Comparison between strong coupling and MC simulation for the
ground state energy in the quadratic model with $V(\varphi)=
\lambda_2\ \varphi^2$ on a lattice with $L=22$ spatial sites.}
\label{fig:SCvsMCE0}
\end{figure}

For large expansion parameter, the strong coupling results can be
compared with explicit simulations (that we shall fully discuss in
Sec.~\ref{Sec:Simulation}).  Let us consider for instance the
quadratic model with $\lambda_0=0$ on a lattice with $L=22$ spatial
sites. In Fig.~\ref{fig:SCvsMCphi}, we show the expectation value
$\langle\varphi_n\rangle$ computed at $\lambda_2=2$. The agreement is
quite good apart from the points on the border where the convergence
seems to be slower. To check the validity of the strong coupling
expansion at smaller couplings, we show in Fig.~\ref{fig:SCvsMCE0} the
ground state energy from MC simulation compared with the first and
second order strong coupling expansion. The scaled variables on the
plot axes are discussed in App.~(\ref{App:Second}). The second order
gives better results at large values of the expansion parameter, but
is unreliable below $\lambda_2\simeq 0.35$.

In the next Section, we shall see that the continuum limit is in the
region of small $\lambda_2$.  Thus, for even $q$, it seems difficult
to gain additional insight from strong coupling and some kind of
transition can happen as the continuum limit is reached.  For this
reason, a full simulation of the model appears to be the only way to
answer the question of symmetry breaking.


\section{Renormalization Group Trajectories}
\label{Sec:RG}

The action of the WZ model is
\[
S = \int d^2x\left(\frac{1}{2} (\partial\varphi)^2 + 
    \frac{i}{2}\overline\psi \partial \psi -
    \frac{1}{2} V(\varphi)^2-\frac{1}{2} V'(\varphi)\overline\psi\psi\right).
\]
At the perturbative level, this is a superrenormalizable field theory
that can be made finite by a renormalization of $V(\varphi)$. In the
minimal subtraction scheme the renormalized potential is obtained by
solving the heat equation~\cite{GFAG}
\begin{eqnarray}
\mu\frac{\partial}{\partial\mu} V(\varphi, \mu) &=& 
-\frac{1}{4\pi} \frac{\partial^2}{\partial\varphi^2}
V(\varphi, \mu) ,
\end{eqnarray}
where $\mu$ is the dimensional regularization scale. A dependence on
$\mu$ is thus introduced in the coefficients of the various monomials
appearing in the tree level $V(\varphi)$. For the specific case of
$V(\varphi) = \lambda_2\varphi^2 + \lambda_0$, we find that
$\lambda_2$ is scale-independent and
\[
\lambda_0(\mu) = \lambda_0(\mu_0)-\frac{\lambda_2}{2\pi}\log\frac{\mu}{\mu_0}.
\]
On the lattice, let us denote by a hat the adimensional lattice
coupling constants and by the label ``ph'' the physical ones, fixed
and with dimension 1. The above result leads to
\begin{eqnarray}
a \lambda_2^{\rm ph} &=& \widehat\lambda_2 \\
a \lambda_0^{\rm ph} &=& \widehat\lambda_0 - 
    \frac{\widehat\lambda_2}{2\pi}\log aM .
\end{eqnarray}
The way we read these equations is: at one loop and for small enough
$a$, the physical $\lambda_0$ is obtained by compensating
$\widehat\lambda_0$ by the effect of the one-loop diagrams.  These are
computed with the UV cutoff $a$ and with the IR cutoff given by the
(dimension 1) mass $M$ of the virtual particles in the loop.

The first equation allows to replace $a$ by $\widehat\lambda_2$
everywhere and we get
\begin{eqnarray}
\widehat\lambda_2 &=& a \lambda_2^{\rm ph} \\
\widehat\lambda_0 &=& \widehat\lambda_2\ 
\frac{\lambda_0^{\rm ph}}{\lambda_2^{\rm ph}} + 
\frac{\widehat\lambda_2}{2\pi}
\log\left(\widehat\lambda_2\frac{M}{\lambda_2^{\rm ph}} \right)
\end{eqnarray}
This seems to show that the continuum limit can be reached with
$\widehat\lambda_2\to 0$ and
\begin{equation}
\frac{\widehat\lambda_0}{\widehat\lambda_2} 
\stackrel{\widehat\lambda_2\to 0}{\sim} 
A +\frac{1}{2\pi}\log\widehat\lambda_2
\label{eq:traj-latt}
\end{equation}
where $A$ contains the ratio $\lambda_2^{\rm ph}/\lambda_0^{\rm ph}$
and the details of the physical mass generation.


\section{Simulation algorithm}
\label{Sec:Simulation}

\subsection{Green Function Monte Carlo: general considerations}

In this Section, we review the Green Function Monte Carlo approach to
the study of the ground state of a general quantum model. To this aim,
we consider the simple case of $0+1$ dimensional quantum mechanics in
order to illustrate the basic ideas without unnecessary details hiding
the essential features of the algorithm.

For a canonical spinless quantum particle, the Hamiltonian is 
\begin{equation}
\label{canonical}
H = \frac 1 2 p^2+V(q),\qquad [q_i, p_j] = i\delta_{i,j} .
\end{equation}
The ground state $|\Psi_0\rangle$ of $H$ can be projected out of any initial 
state $|i\rangle$ with  non zero overlap $\langle \Psi_0 | i \rangle
\neq 0$. The projection is performed by applying the evolution semigroup 
$\{\exp(-t H)\}_{t\ge 0}$ and going to asymptotically large times.

Focusing on the ground state energy $E_0$, this procedure leads to the
following simple formula
\begin{equation}
E_0 = \lim_{t\to +\infty}
\frac{\langle f| H e^{-t H} | i\rangle} {\langle f| e^{-t H} |
i\rangle};
\label{eq:E0}
\end{equation}
the final state $|f\rangle$ is in principle arbitrary, as long as it
is not orthogonal to $|\Psi_0$; in practice, it must
be chosen with care, to avoid numerical instability.

The vacuum expectation value of a generic observable $O$ can be computed
as
\begin{equation}
\langle \Psi_0 | O | \Psi_0 \rangle =
\lim_{t,\tau\to\infty} 
    {\langle f|e^{-\tau H}\,O\,e^{-t H}|i\rangle\over
     \langle f|e^{-(\tau+t) H}|i\rangle};
\label{eq:forward-walking}
\end{equation}
this procedure is known as {\em forward walking}.

To translate the above formula into a stable numerical algorithm, it
is necessary to find a basis such that the Hamiltonian $H$ has non
positive off-diagonal matrix elements. By the way, this is true for
the Hamiltonian~(\ref{canonical}) in the basis $\{|q\rangle\}$ of
position eigenstates.  If such a basis is found, it is possible to
identify matrix elements of $e^{-t H}$ as probability transitions
defining a Markov random process in the state space. For instance, in
the simple case when $|f\rangle$ is chosen to be a zero momentum
state, $p |f\rangle = 0$, we have (Feynman-Kac formula)
\begin{equation}
\label{fk}
E_0 = \lim_{t\to +\infty}
\frac{\langle f| V e^{-t H} | i\rangle} {\langle f| e^{-t H} |
i\rangle} = \lim_{t\to +\infty}
\frac{\displaystyle 
\int{\cal D}q(\tau)\ V(q(t)) e^{-\int_0^t V(q(\tau)) d\tau}}
{\int{\cal D}q(\tau)\ e^{-\int_0^t V(q(\tau)) d\tau}},
\end{equation}
where ${\cal D}q(\tau)$ is the Wiener measure.

The probabilistic interpretation of the above equation is as follows:
$E_0$ (as well as other observables) can be computed 
by taking the average over weighted walkers which diffuse
according to the Wiener process. Each path is weighted by the 
following functional of the trajectory
\begin{equation}
\label{weight}
W[q(\tau)] = \exp -\int_0^t V(q(\tau)) d\tau .
\end{equation}

In the numerical implementation, an 
estimate of $E_0$ is obtained by computing
\begin{equation}
\lim_{t\to +\infty} 
\frac{\mbox{\bf E}\left(V(\widehat{q}_t)\widehat{W}_t\right)}%
{\mbox{\bf E}\left(\widehat{W}_t\right)},
\end{equation}
where $\widehat{q}_t$ is a numerical discretization of the Wiener
process, $\widehat{W}_t$ its associated weight computed by properly
approximating Eq.~(\ref{weight}) and, finally, $\mbox{\bf
E}\left(\cdot\right)$ denotes the average with respect to the
realizations of $\widehat{q}_t$. In practice, after the choice of a
particular approximation $\widehat{q}_t$, one works with a large
number $K$ of walkers and extrapolates numerically to $K\to \infty$.
The control of the approximations involved in this strategy requires
some discussion that we defer to the Section devoted to results.

A point that is worth mentioning regards the possibility of
introducing a guidance in the walkers diffusion. To improve the
convergence to ground state it is customary to define the unitarily
equivalent Hamiltonian
\begin{equation}
\label{unitary}
\widetilde H = e^S H e^{-S} = 
\frac 1 2 p^2 + i p \cdot F + \widetilde V(q) , 
\end{equation}
where $S$ is an arbitrary (real) function and 
\begin{eqnarray}
\label{impsamp}
F &=& \nabla S , \\
\widetilde V &=& V-\frac 1 2 (\nabla S)^2 -\frac 1 2\Delta S .
\nonumber 
\end{eqnarray}
It can be shown that the derivation of expressions like~(\ref{fk}) can
be easily generalized to this case and the required modifications can
be summarized as: (i) the potential $V$ is replaced by with
$\widetilde V$, (ii) the Wiener process is replaced by a deformed
process guided by the drift $F$.  In the following, we shall call
Importance Sampling the trick of exploiting a non zero $S$.

In the following Sections, we describe in full details 
the algorithm for the model under study considering first
the bosonic and fermionic sectors separately and finally the full Hamiltonian.

\subsection{Bosonic sector}
\label{subsec:bosalg}

The bosonic sector of the lattice model is a canonical quantum
mechanical one with many degrees of freedom. The algorithm is the one
described in the previous Section. Given the transformed Hamiltonian
$\widetilde H$ of Eq.~(\ref{unitary}), we write
\begin{equation}
\label{factorization}
\exp(-\varepsilon \widetilde H_B) = 
\exp\left(-\frac{\varepsilon}{2} \widetilde V\right) 
\exp\left(-\frac{\varepsilon}{2}\frac{p^2}{2}\right)
e^{- i \varepsilon p\cdot F}
\exp\left(-\frac{\varepsilon}{2}\frac{p^2}{2}\right)
\exp\left(-\frac{\varepsilon}{2}\widetilde V\right) + {\cal O}(\varepsilon^3).
\end{equation}
The function $\widetilde V$ depends on the bosonic state, i.e. the set of 
values of the scalar fields that we collectively denote by $Q$.

The update rule for the weighted walker $(Q, W)$ is built, step by
step, following the approximate operator
factorization~(\ref{factorization}) and reads (see~\cite{Chin} for
similar calculations in the solution of the Langevin equation)
\begin{equation}
(Q', W')\to (Q''', W'''),
\end{equation}
where $Q'''$ and $W'''$ are built according to 
\begin{eqnarray}
W'' &=& W' \exp\left(-\frac \varepsilon 2  \widetilde V(Q')\right), \\
Q'' &=& Q' + \sqrt\frac{\varepsilon}{2} \xi', \nonumber \\
z  &=& Q' + \frac \varepsilon 2 F(Q'), \nonumber\\
Q'' &=& Q' + \varepsilon F(z), \nonumber \\
Q''' &=& Q'' + \sqrt\frac{\varepsilon}{2} \xi'', \nonumber \\
W''' &=& W'' \exp\left(-\frac \varepsilon 2 
\widetilde V(Q''')\right), \nonumber
\end{eqnarray}
and $\xi'$, $\xi''$ are independent sets of Gaussian random numbers.
In the above update, the 
integration of the equations of motion associated with the
driving force $F$ has been solved at second order.
In the end a systematic error ${\cal O}(\varepsilon^3)$ with respect to the
evolution time has been introduced.

An estimate of the energy in the bosonic sector is obtained by taking
the weighted average of $\widetilde V$ over several realizations of 
the walker path
\begin{equation}
E_{0}^{\rm bosonic} = 
\lim_{t\to +\infty}
\frac{\mbox{\bf E}\left(\widetilde V(Q_t)\ W_t\right)}%
{\mbox{\bf E}\left(W_t\right)} .
\end{equation}

\subsection{Fermionic sector}
\label{subsec:feralg}

In the fermionic sector, the spirit of the algorithm is the same, but
there are important technical differences that we want to emphasize.
At fixed scalar fields configuration, the remaining state space is
purely fermionic and, on a finite lattice, it is both discrete and
finite dimensional.

To simplify notation, in this Section we denote $H\equiv H_F$.
The Hamiltonian can be thought as a large sparse matrix $H =
\Vert H_{ss'}\Vert$ with $s$ and $s'$ denoting fermionic states.
We now show that a similar construction like the one exploited 
with $H_B$ can be repeated here.
The Gaussian random noise that was the building block in the simulation
of the Wiener process is replaced here by a discrete jump 
process.

Again, the problem is that of giving a probabilistic
representation for the evolution semigroup $\Omega=\{e^{-tH}\}_{t\ge 0}$.
To this aim, we define a Markov process that describes diffusion in
the discrete state space and also provide a rule for the update of a
walker weight. We finally show that suitable averages over weighted
walkers reconstruct the evolution governed by $H$ and project
a given initial state onto the ground state.

For each pair $s, s'$ in the state space $S$ 
such that $s\neq s'$ and $H_{s's}\neq 0$
we define $\Gamma_{s's} = -H_{s's}$. We assume that all
$\Gamma_{s's}>0$ (no sign problem)
and build a $S$-valued Markov stochastic process $s_t$
by identifying $\Gamma_{s's}$ as the rate for
the transition $s\to s'$. Hence, 
the average occupation 
$
P_s(t) = \mbox{\bf E}\left(\delta_{s, s_t}\right),
$
with $\mbox{\bf E}\left(\cdot\right)$ denoting the average with respect 
to $s_t$, obeys the Master Equation 
$
\dot P_s(
\beta) = \sum_{s'\neq s}(\Gamma_{ss'}
P_{s'}-\Gamma_{s's} P_s) .  
$

Related to $s_t$, we also define the real valued stochastic process
$W_t = \exp\left(-\int_0^t \omega_{s_t}\ dt\right)$, with 
$\omega_s = \sum_{s'\in S} H_{s's}$.  It can be shown that the
weighted expectation value 
$\psi_s(t) = \mbox{\bf E}\left(\delta_{s, s_t} W_t\right)$
reconstructs $\Omega$:
\[
\frac{d}{dt}\psi_s(t) = -\sum_{s'\in S} H_{s s'}\psi_{s'}(t) ,
\]
with $\psi_s(0) = \mbox{Prob}(s_0=s)$. 
Matrix elements of $\Omega$
can be identified with certain expectation values. 
In particular, the ground state
energy $E_0$ (in the purely fermionic sector) can be obtained by 
\begin{equation}
E_0 = \lim_{t\to +\infty} 
\frac{\mbox{\bf E}\left(\omega_{s_t}\ W_t\right)}%
{\mbox{\bf E}\left(W_t\right)} .
\end{equation}
The actual construction of the process is straightforward.
A realization of $s_t$
 is a piece-wise constant map ${\bf R}\to S$
with isolated jumps at times $t=t_0, t_1, \dots$, with
$t_0<t_1<t_2<\cdots$. 
The algorithm that computes the triples $\{t_n, s_{t_n},
W_{t_n}\}$ is the following:
\begin{enumerate}

\item We  denote $s_{t_n} \equiv s$ and 
define the set $T_s$ of target states connected to $s$:
$T_s = \{s', \Gamma_{s's}>0\}$. We also define the 
total width $\Gamma_s = \sum_{s'\in T_s} \Gamma_{s's}$.

\item Extract $\tau\ge 0$ with probability density 
$p_s(\tau) = \Gamma_s e^{-\Gamma_s\tau}$. 
In other words, $\tau = -\frac 1 {\Gamma_s} \log\xi$ with $\xi$ 
uniformly distributed in $[0,1]$. 

\item Extract a new state $s'\in T_s$ with probability
$p_{s'} = \Gamma_{s's}/\Gamma_s$.

\item Define $t_{n+1} = t_n + \tau$, $s_{t_{n+1}} = s'$ and 
$W_{t_{n+1}} = W_{t_n}\cdot e^{-\omega_s\tau}$.
\end{enumerate}
In this sector there is no systematic error due to a
finite evolution time. The semigroup dynamics is in fact reproduced
exactly by the above stochastic process $(s_t, W_t)$.

About Importance Sampling, we remark that in the discrete case, 
the inclusion of a trial wave function amounts to the redefinition 
\begin{equation}
\label{trialfermion}
\widetilde H_{s's} = \Psi_{s'}^T H_{s's} \frac{1}{\Psi_s^T} , 
\end{equation}
where $\Psi_s^T = \langle s | \Psi_0^T\rangle$ are the components of the
trial ground state $|\Psi_0^T\rangle$. The new Hamiltonian $\widetilde
H$ is not symmetric, but the above formulas works as well with no need 
for further modifications: actually, they have been derived without requiring 
any symmetry condition $H = H^T$.

Some final comments are in order about the choice of the basis for the
fermionic states.  As we mentioned in the general discussion, we want
to have zero or negative off-diagonal matrix elements of $H$. The
simplest choice amounts to consider eigenstates $|{\bm n}\rangle$ of
the occupation numbers $\chi^\dagger_i\chi_i$ and with a relative
phase fixed by the natural choice
\begin{equation}
|{\bm n}\rangle = \prod_{i=1}^L (\chi^\dagger)^{n_i} |0\rangle,\qquad 
\chi_i|0\rangle = 0 .
\end{equation}
This does not guarantee that the above sign problems are absent. In
fact, in weak-coupling perturbation theory, the choice of periodic
boundary conditions does not break supersymmetry when $L\mathop{\rm
mod}4 = 0$ as can be checked, e.g., in the free model.  However, under
this condition, there is an even number of fermions, $L/2$, in the
ground state and a sign problem arises due to boundary crossings of a
fermion, since such a transition involves an odd number of fermion
exchanges.  To avoid such a difficulty, we shall adopt open boundary
conditions.  With this choice, $L$ needs just to be even to assure a
supersymmetric weak coupling ground state.  Also, we shall restrict to
the case $L\mathop{\rm mod}4 = 2$ and to the sector with $L/2$
fermions that contains a non-degenerate ground state, with zero energy
at all orders in a weak-coupling expansion.

\subsection{Algorithm for the full model}

To study the full Hamiltonian of the Wess Zumino model, the simplest
attitude is to perform an approximate splitting of the bosonic and 
fermionic sectors. For instance, with second order precision,
we can write
\begin{equation}
\label{splitting}
\exp\left(-\varepsilon H\right) = 
\exp\left(-\frac 1 2\ \varepsilon H_B\right) 
\exp\left(-\varepsilon H_F\right) 
\exp\left(-\frac 1 2 \ \varepsilon H_B\right)+ {\cal O}(\varepsilon^3) , 
\end{equation}
and consider separately the evolution related to $H_B$ and $H_F$
freezing the fermionic or bosonic fields respectively. In the end, an
extrapolation to the $\varepsilon\to 0$ limit must be performed.
Eq.~(\ref{splitting}) has been approximated to the same order as
Eq.~(\ref{factorization}); if necessary, both can be improved.

\subsection{Variance control}
\label{subsec:sr}

A straightforward implementation of the above formulae fails because
of a numerical instability: the variance of the walker weights $W_t$
computed over the walkers ensemble grows exponentially with $t$ and
forbids the projection onto the ground state \cite{Hetherington}.  A
good trial wave function can certainly reduce the growth rate, but the
problem disappears only in the ideal case when the trial wave function
is exact.  To bypass this problem, some kind of {\em branching\/} procedure
must be applied in order to delete trajectories (walkers) with low
weight and replicate those with larger weight.

In practice, we introduce an ensemble, i.e., a collection of $K$
independent walkers $s^{(n)}$ each one carrying its own weight
$W^{(n)}$:
\begin{equation} 
\ens = \{(s^{(n)}(t), W^{(n)}(t)), 1\le n\le K \}.
\end{equation}
When the variance of the weights in the ensemble becomes too large,
$\ens$ is transformed in a new ensemble $\ens'$ that reproduces the
same expectation values (at least in the $K\to\infty$ limit) and has a
smaller variance.  We adopted the branching procedure of Ref.\ 
\cite{Trivedi-Ceperley-90}: for each walker $s^{(n)}$ we compute a
multiplicity
\begin{equation}
M^{(n)} = \lfloor c W^{(n)} + \xi\rfloor, \qquad
c = {{\bar K}\over\sum_n W^{(n)}},
\end{equation}
where $\xi$ is a random number uniformly distributed in $[0,1]$,
${\bar K}$ is the desired number of walkers, and $\lfloor x\rfloor$ is
the maximum integer not greater than $x$; the new ensemble $\ens'$
contains $M^{(n)}$ copies of each configuration $s^{(n)}$ in $\ens$
and all the weights are set to $1$; the actual number of walkers $K$
will oscillate around ${\bar K}$.  This procedure has the advantage
that there is no harmful effect from its repeated application;
therefore we apply it at each Monte Carlo iteration, after all the
fields have been updated.

\subsection{Choice and dynamical optimization 
            of the trial wave function $\bm{\Psi_T}$}
\label{Sec:trial-WF}

About the choice of the trial wave function, we propose 
the factorized form
\begin{equation}
|\Psi_0^T\rangle = e^{S_B(\varphi)+S_F(\varphi, \chi,
\chi^\dagger)} | \Psi_0\rangle_B \otimes
|\Psi_0\rangle_F, 
\end{equation}
where $|\Psi_0\rangle_B \otimes
|\Psi_0\rangle_F$ is the ground state of the free model given explicitely
in App.~(\ref{App:Free}) and
\[
S_B = \sum_n \sum_{k= 1}^{d_B} \alpha^B_k \varphi_n^k, \quad S_F =
\sum_n (-1)^n \left(\chi_n^\dagger\chi_n-\frac 1 2\right)\
\sum_{k=1}^{d_F} \alpha^F_k \varphi_n^k.
\]
Since the trial wave function is a modification of the free
ground-state wave function, we expect that importance sampling
will improve as we approach the continuum limit.

The degrees $d_B$ and $d_F$ must be chosen carefully to achieve a
balance between the accuracy of the trial wave function on one hand,
and convergence of the adaptive determination of the parameters
$\alpha$ and computation time on the other hand.  We chose
$d_B=d_F=4$, except for situations very close to the continuum limit,
e.g., $V=\lambda_2\varphi^2+\lambda_0$ with $\lambda_2<0.2$, for which we
chose $d_B=d_F=2$ (cf.\ Sect.\ \ref{Sec:continuum}).

The trial wave function should of course respect the symmetries of the
model; a $Z_2$ symmetry possessed by the model for specific forms of
$V$ is very helpful to reduce the number of parameters that we must
optimize.  For odd $V$, the model enjoys the exact symmetry
$\varphi_n\to-\varphi_n$, and therefore odd $\alpha^B_k$ and $\alpha^F_k$
can be set to zero.  For even $V$, the model enjoys the approximate
symmetry $\varphi_n\to-\varphi_n$, $\chi_n\leftrightarrow\chi_n^\dagger$ (it
is broken by irrelevant terms and by boundary terms), and we verified
that odd $\alpha^B_k$ and even $\alpha^F_k$ can be set to zero.

Let us denote by $\boa = \{\alpha^B, \alpha^F\}$ the collective set of
free parameters appearing in the trial wave function. A possible
approach consists in performing simulations of moderate size at fixed
$\boa$ in order to optimize their choice.  However, as shown
in~\cite{MB}, the trial wave function $|\Psi_0^T\rangle$ can also be
optimized dynamically within Monte Carlo evolution with a better
performance of the full procedure.

The idea is again simple: consider the ground state energy as a
typical observable; for a given choice of $\boa$, after $N$ Monte
Carlo steps, a simulation with an average population of $K$ walkers
furnishes a biased estimator $\hat E_0(N, K, \boa)$.  If we denote by
$\langle\cdot\rangle$ the average with respect to Monte Carlo
realizations, $\hat E_0$ is a random variable such that
\begin{equation}
\lim_{N\to\infty} 
\langle \hat
E_0(N, K, \boa) \rangle = E_0 + \delta E_0(\boa, K),
\end{equation}
where $\delta E_0(\boa, K)$ depends on $\boa$, but vanishes as $K\to \infty$. 
Besides, the size of the fluctuations is measured by 
\begin{equation}
\mbox{Var}\ \hat E_0(N, K, \boa) \stackrel{N\to\infty}{\sim} 
\frac{c_2(K, \boa)}{\sqrt{N}}.
\end{equation}
In the $K\to\infty$ limit, $\langle \hat E_0\rangle$ is exact and
independent on $\boa$.  The constant $c_2(K, \boa)$ is related to the
fluctuations of the effective potential $\widetilde V$ and is strongly
dependent on $\boa$. The problem of finding the optimal trial wave
function can be translated in the minimization of $c_2(K, \boa)$ with
respect to $\boa$.

The algorithm we propose
performs this task by interlacing a Stochastic
Gradient steepest descent with the Monte Carlo evolution of the walkers
ensemble. At each Monte Carlo step, we update
$\boa_n\to\boa_{n+1}$ according to the simple law 
\begin{equation}
\boa_{n+1} =
\boa_n -\eta_n \nabla_\boa \mbox{Var}_{\ens_n}\ \widetilde V 
\end{equation}
where ${\cal E}_n$ is the ensemble at step $n$ and $\{\eta\}$ is a
suitable sequence, asymptotically vanishing; to keep things simple, we
use the same $\eta$ for all components $\alpha_k$ of $\boa$, although
in principle we could use a different $\eta_k$ for each $\alpha_k$.

A non-linear feedback is thus established between the trial wave
function and the evolution of the walkers. The convergence of the
method can not be easily investigated by analytical methods and
explicit numerical simulations are required to understand the
stability of the algorithm.  In~\cite{MB}, examples of applications of
the method with purely bosonic or fermionic degrees of freedom can be
found.  Here, we apply the method for the first time to a model with
both kinds of fields.

In practice, the choice of the initial values of $\boa$ is important:
it is clear that, if we have a good guess of the optimal values (e.g.,
from runs at the same $V$ but for smaller values of $L$ or $K$),
starting from them makes the convergence much faster.  We also noticed
that, starting e.g.\ from $\boa_0=0$, the steepest descent at times
fails and $\boa_n$ oscillates wildly; this never happens if most of
the starting values have at least the right sign and order of
magnitude.

We found it useful to determine $\eta$ dynamically as well.  The basic
idea is that we wish to decrease $\eta$ when all the $\alpha_k$ have
reached the optimal value and they are oscillating at random around
it: in this situation, a smaller $\eta$ means less noise on $\boa$.
On the other hand, we wish to increase $\eta$ when one or more
$\alpha_k$ is drifting: a larger $\eta$ now means a faster approach
toward the optimal value.  To monitor the trend of $\boa$, we use the
quantity
\begin{equation}
\tau = \max_k \tau_k, \qquad 
\tau_k =  {N |a_k| \over v_k} - 3,
\end{equation}
where $N=n_1-n_0$ is the number of iterations in the interval of Monte
Carlo iterations $(n_0,n_1]$ we are considering, $v_k$ is the variance
of $\alpha_k$ in the interval, and $a_k$ is the slope of the linear
least-squares fit to $\alpha_{k,n}$ vs.\ $n$.  $\tau_k$ is invariant
under translations and scale transformations; it is positive if
$\alpha_k$ is drifting and it is negative if $\alpha_k$ is
oscillating.

\begin{figure}[tbp]
\begin{center}
\leavevmode
\epsfig{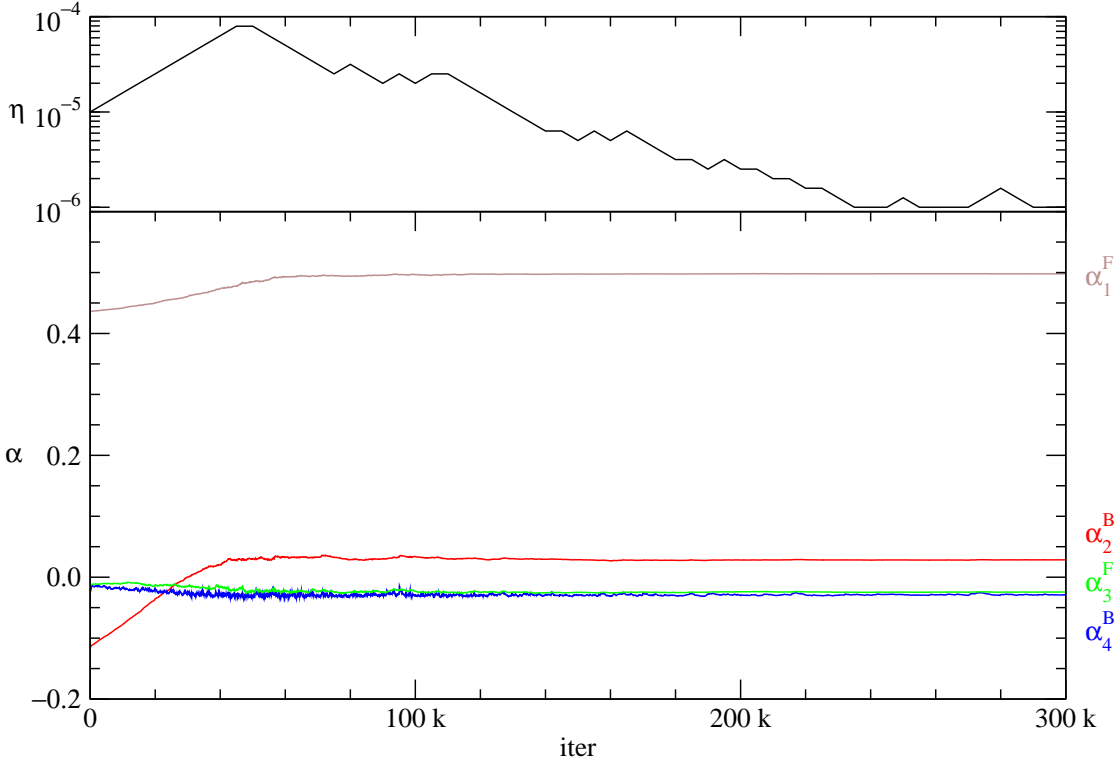}
\vskip-7mm
\end{center}
\caption{$\boa$ and $\eta$ dynamics from a run at
$V = 0.5\,\varphi^2 - 0.55$, $L=34$ and $K=100$.}
\label{fig:alpha+eta}
\end{figure}

A typical example of the implementation of the $\boa$ and $\eta$
dynamics is shown in Fig.\ \ref{fig:alpha+eta}; $\eta$ is initialized
to $10^{-5}$; after each interval of $N=5000$ iterations, if
$\tau>0.15$, $\eta$ is multiplied by $\root10\of{10}$; if $\tau<0$,
$\eta$ is divided by $\root10\of{10}$; if $0<\tau<0.15$, $\eta$ is
unchanged; finally, $\eta$ is restricted to the interval
$[10^{-6},10^{-4}]$.  The parameters of the $\eta$ dynamics given here
were obtained empirically.

\subsection{Observable measurement}

We measure the vacuum expectation values of $\varphi_n$, $\varphi_n\varphi_m$,
$\chi^\dagger_n\chi_n$, and of
\begin{equation}
T = \half\sum_{n=1}^L (\varphi_{n+1}-\varphi_{n-1}) \, V(\varphi_n), \qquad
Y_q = \Bigl\{Q,\ \sum_{n=1}^L \varphi_n^q \psi_{2,n}\Bigr\}. 
\label{eq:observ}
\end{equation}
Note that, with our choice of boundary conditions, we don't have
translation invariance and, e.g., $\langle\varphi_n\rangle$ will depend
on $i$; however, the dependence is sizable only within a few
correlation lengths from the border; therefore we typically average
site-dependent quantities excluding sites closer than a suitable
$L_{\rm min}$ from the border; in the case of
$\langle\varphi_n\varphi_m\rangle$, we average over all pairs with fixed
distance $r=|n-m|$, excluding the cases when $n$ or $m$ is closer than
$L_{\rm min}$ from the border.

\begin{figure}[tbp]
\begin{center}
\leavevmode
\epsfig{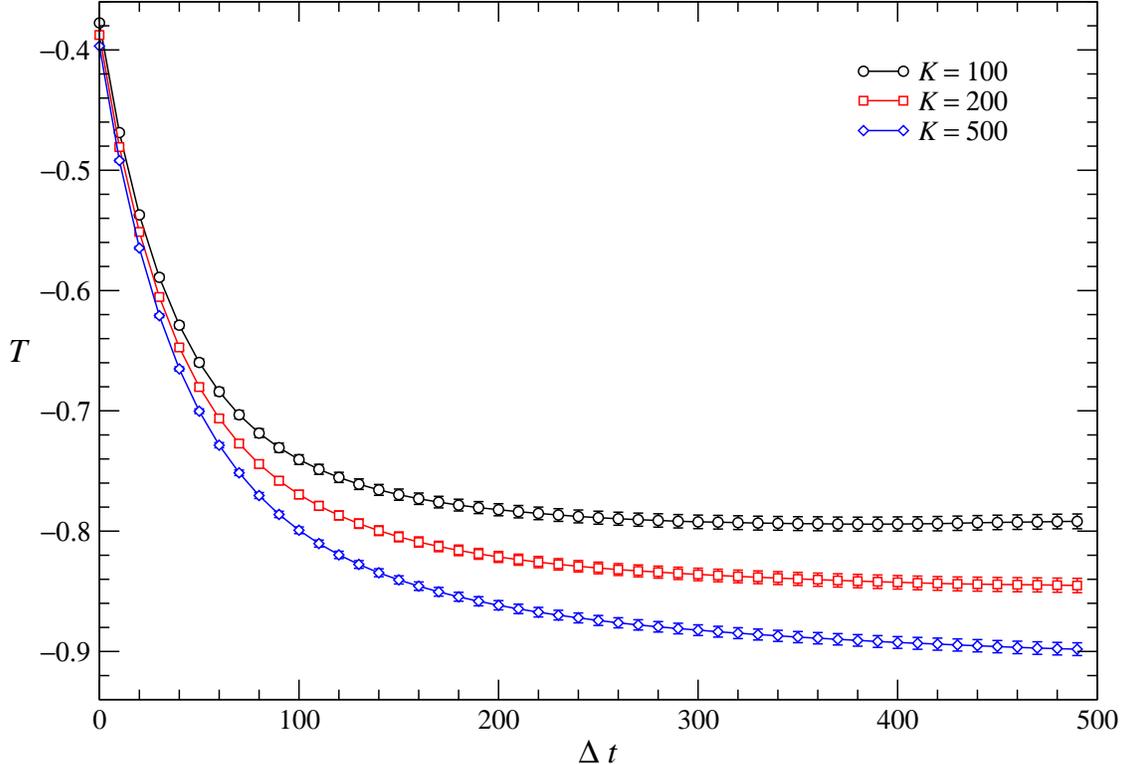}
\vskip-7mm
\end{center}
\caption{The central charge $T$ (cf.\ Eq.\ (\ref{eq:observ}))
vs.\ $\Delta t$ from runs at $V = 0.5\,\varphi^2$ and $L=34$.}
\label{fig:Deltat}
\end{figure}

The ground-state energy is measured simply by averaging the measured
values of $E_0$ over the ensembles $\ens(t)$, discarding a suitable
thermalization interval $(0,t_0)$:
\begin{equation}
E_0 \cong {1\over t_1-t_0}\sum_{t=t_0+1}^{t_1} 
{\textstyle\sum_{i=1}^{K_t} E_{i,t} w_{i,t}
 \over\textstyle\sum_{i=1}^{K_t} w_{i,t}}, \qquad
E_{i,t} = 
{\langle s_{i,t}|H|s_{i,t}\rangle \over \langle s_{i,t}|s_{i,t}\rangle},
\end{equation}
cf.\ Eq.\ (\ref{eq:E0}).  The vacuum expectation value of a generic
observable is computed implementing the forward-walking formula
(\ref{eq:forward-walking}) as
\begin{equation}
\langle O\rangle \cong {1\over t_1-t_0}\sum_{t=t_0+1}^{t_1} 
{\textstyle\sum_{i=1}^{K_t} O_{i,t} w_{i,t+\Delta t}
 \over\textstyle\sum_{i=1}^{K_t} w_{i,t+\Delta t}}, \qquad
O_{i,t} = 
{\langle s_{i,t}|O|s_{i,t}\rangle \over \langle s_{i,t}|s_{i,t}\rangle};
\label{eq:O}
\end{equation}
In principle, in Eq.\ (\ref{eq:O}) we must take the 
$\Delta t\to\infty$ limit, but in practice a moderate value like
$\Delta t = 500$ is sufficient.  A typical examples of $\Delta t$
dependence is shown in Fig.\ \ref{fig:Deltat}; it should be noticed
that the error bars grow with $\Delta t$ but very slowly.

\section{Numerical results}
\label{Sec:Results}

\subsection{Review of previous lattice studies}

The class of models that we study in this paper has been previously considered
in~\cite{Ranft-Schiller} with a Monte Carlo approach that determine
the ground state energy by using
\begin{equation}
E_0  = \lim_{\beta\to\infty}
\frac{\mbox{Tr}(H e^{-\beta H})}{\mbox{Tr}(e^{-\beta H})} ,
\end{equation}
and working numerically with a large fixed $\beta$. This is in the
spirit of the usual Lagrangian algorithms to be compared with the
Green Function Monte Carlo method where $\beta$ can be identified with
the simulation time and is thus taken to infinity by the very nature
of the algorithm.

The analysis of~\cite{Ranft-Schiller} is performed on $12\times 100$
lattice, hence with a rather coarse spatial mesh. In the model with
$V(\varphi) = \lambda_3\varphi^3$ supersymmetry appears to be unbroken
in full agreement with our analysis. In the quadratic model with
$V(\varphi) = \lambda_2\varphi^2+\lambda_0$, the authors of Ref.\ 
\cite{Ranft-Schiller} find rather strong signals for supersymmetry
breaking with $\lambda_0$ bigger that the critical value $\lambda_0
\simeq -0.5$ and have numerical results showing a very small ground
state energy for $\lambda_0 < -0.5$.  No discussion of the continuum
limit is attempted.

\subsection{Odd $\bm V$}

\begin{figure}[tbp]
\begin{center}
\leavevmode
\epsfig{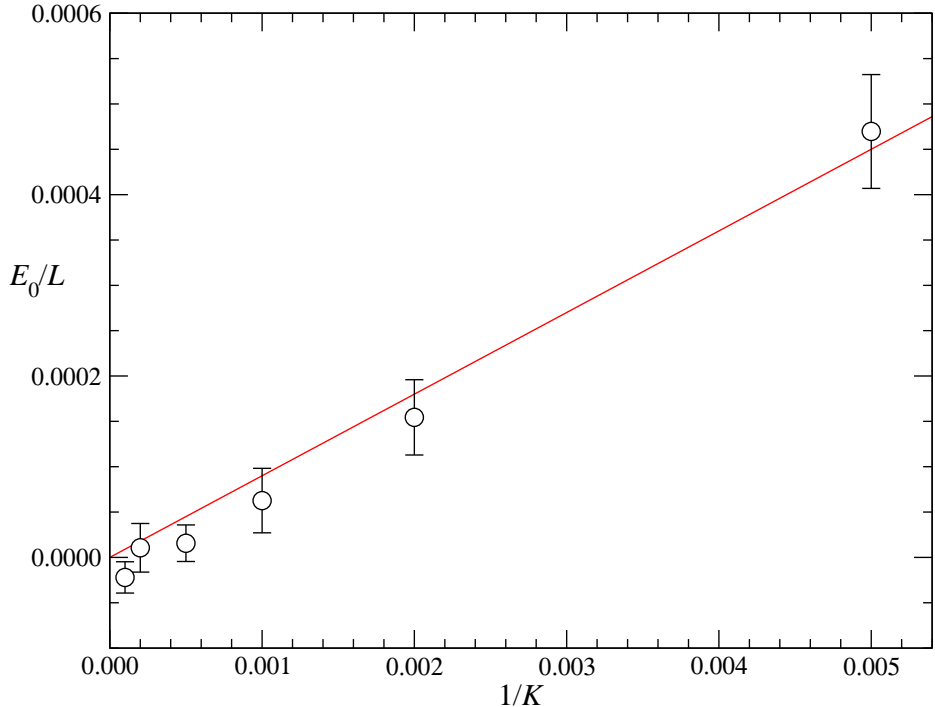}
\vskip-7mm
\end{center}
\caption{The ground-state energy density $E_0/L$ vs.\
$1/K$ at $V = \varphi^3$, $L=22$, with statistics of 1~{\rm M} iterations
for $K<5000$, 500~{\rm k} iterations at $K=5000$, and 
300~{\rm k} iterations at $K=10000$.}
\label{fig:Cubic-E0}
\end{figure}

\begin{figure}[tbp]
\begin{center}
\leavevmode
\epsfig{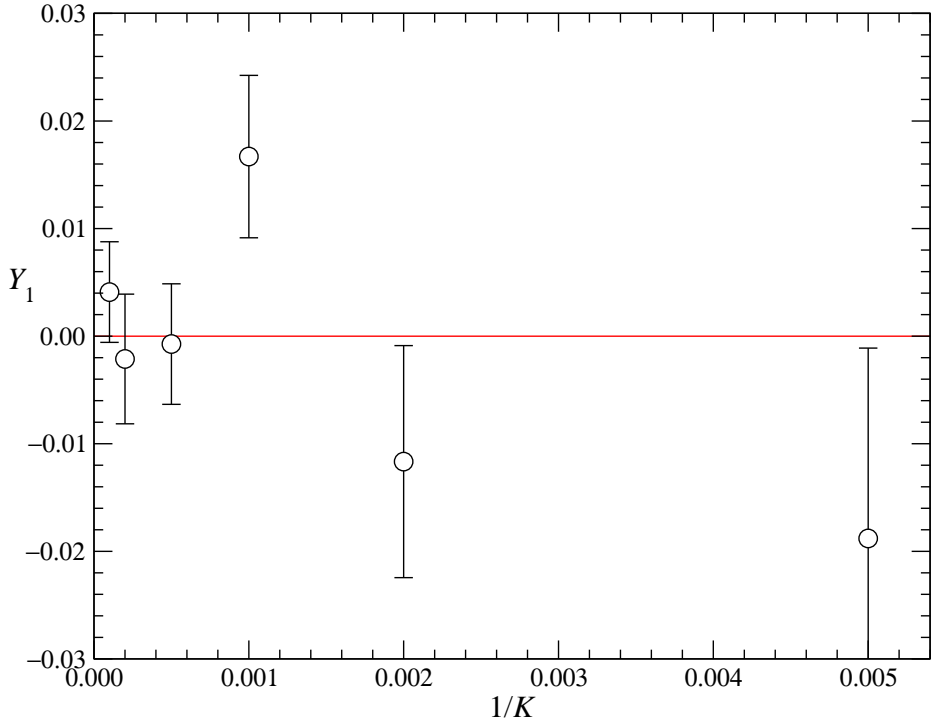}
\vskip-7mm
\end{center}
\caption{The Ward identity $Y_1$ vs.\ $1/K$ at $V = \varphi^3$, $L=22$.}
\label{fig:Cubic-Y1}
\end{figure}

As an example of odd prepotential, we study the case $V=\varphi^3$.  We
plot the ground-state energy vs.\ $K$ in Fig.\ \ref{fig:Cubic-E0} and
the Ward identity $Y_1$ vs.\ $K$ in Fig.\ \ref{fig:Cubic-Y1}.  Both
give a very convincing evidence for unbroken SUSY; all the other Ward
identities are consistent with zero, but more noisy.  It should be
noticed that the bosonic and fermionic contribution to $E_0$ are
$\simeq \pm7.4$: we are observing a cancellation of four orders of
magnitude.

\subsection{Supersymmetry breaking}
\label{Sec:SSB}

We now turn to the more interesting case of even prepotential,
investigating the case $V=\lambda_2\varphi^2 + \lambda_0$.  We remind
that in this case the model enjoys an approximate $Z_2$ symmetry
$\varphi_n\to-\varphi_n$, $\chi_n\leftrightarrow\chi_n^\dagger$.  For fixed
$\lambda_2$, we may expect (in the $L\to\infty$ limit) a phase transition at
$\lambda_0=\lambda_0^{(c)}(\lambda_2)$, separating a phase of broken
SUSY and unbroken $Z_2$ (high $\lambda_0$) from a phase of unbroken
SUSY and broken $Z_2$ (low $\lambda_0$).  We investigated in details
the case $\lambda_2=0.5$.

The usual technique for the study of a phase transition is the
{\em crossing\/} method, applied to the Binder cumulant \cite{Binder}
\begin{equation}
B = {1\over2}\left(3 - {\langle M^4\rangle\over\langle M^2\rangle^2}\right);
\label{eq:B}
\end{equation}
in our case, the choice of a sensible definition for the magnetization
$M$ is nontrivial, since our model is neither ferromagnetic nor
antiferronagnetic, and it doesn't enjoy translation symmetry.  We
tried out several definitions, before making the choice
\begin{equation}
M_{\rm even\,(odd)} \equiv {2\over L - 2 L_{\rm min}}
\sum_{{\rm even\,(odd)}\,i=1+L_{\rm min}}^{L-L_{\rm min}} \phi_i,
\label{eq:M}
\end{equation}
where, typically, $L_{\rm min}=6$; $M_{\rm even}$ and $M_{\rm odd}$
are perfectly equivalent, and the values of $B$ we quote in the
following are the average of $B_{\rm even}$ and $B_{\rm odd}$.

\begin{figure}[tbp]
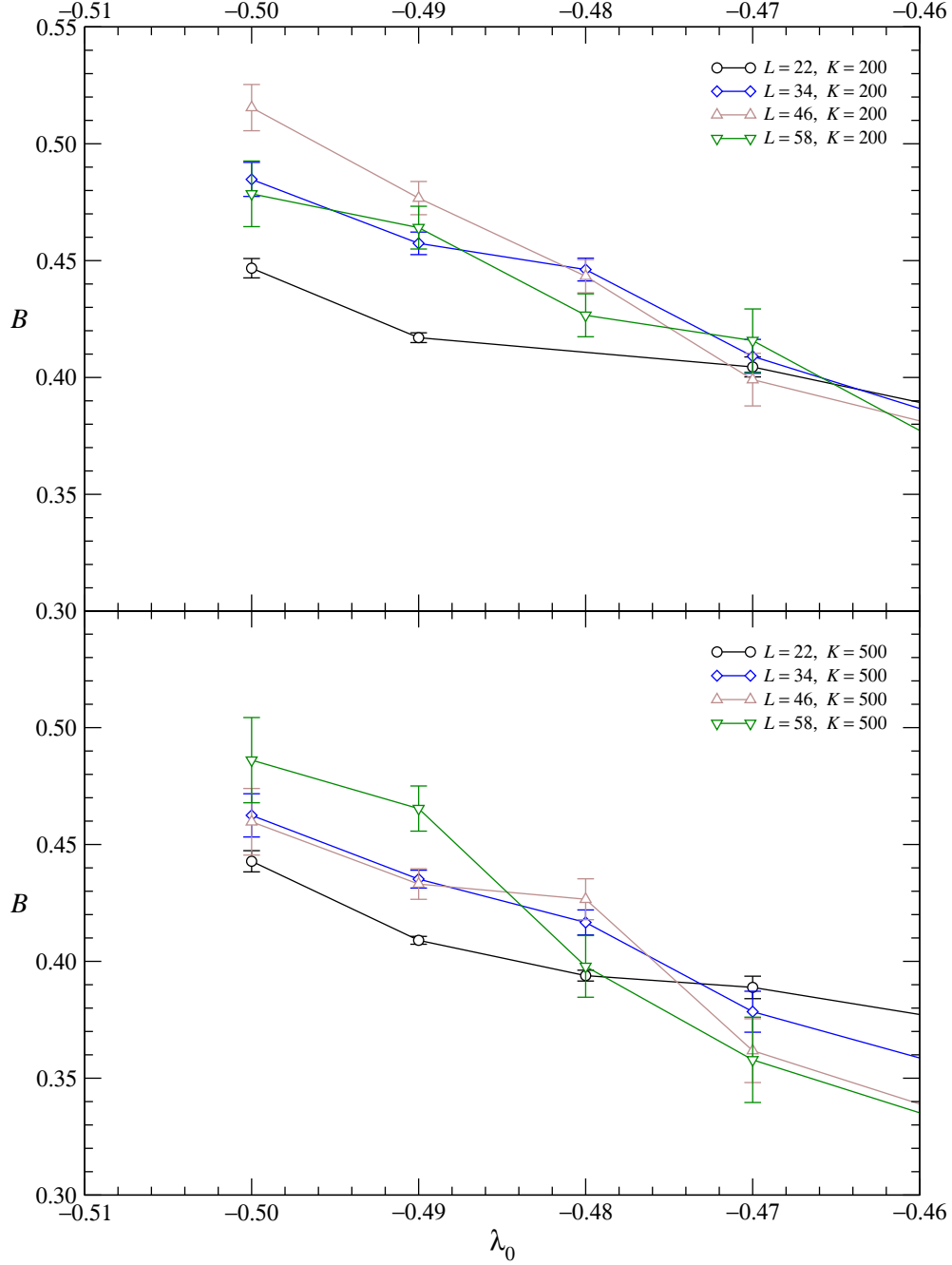

\begin{center}
\leavevmode
\epsfig{file=B.K_200.eps,width=0.8\textwidth,angle=0}
\vskip-2.3mm
\epsfig{file=B.K_500.eps,width=0.8\textwidth,angle=0}
\vskip-7mm
\end{center}
\caption{The Binder cumulant $B$ vs. $\lambda_0$.}
\label{fig:Binder}
\end{figure}

The crossing method consists in plotting $B$ vs.\ $\lambda_0$ for
several values of $L$; the crossing point $\lambda_0^{\rm
cr}(L_1,L_2)$, determined by the condition
\[
B(\lambda_0^{\rm cr},L_1) = B(\lambda_0^{\rm cr},L_2),
\]
is an estimator of $\lambda_0^{(c)}$ \cite{Binder}; the convergence is
dominated by the critical exponent $\nu$ of the  correlation length and by the
critical exponent $\omega$ of the leading corrections to scaling
(cf.\ Ref.\ \cite{PV-rept}):
\[
\lambda_0^{\rm cr}(L_1,L_2) = \lambda_0^{(c)} + 
O(L_1^{-\omega-1/\nu},L_2^{-\omega-1/\nu});
\]
we expect the phase transition we are studying to be in the Ising
universality class, for which $\nu=1$ and $\omega=2$, and therefore we
expect fast convergence of $\lambda_0^{\rm cr}$ to $\lambda_0^{(c)}$.
The results, plotted in Fig.\ \ref{fig:Binder}, indicate
$\lambda_0^{(c)}=-0.48\pm0.01$.

It is possible to study the phase transition by looking at the
connected correlation function
$G_d=\langle\varphi_n\varphi_{m}\rangle_c$, averaged over all $n,m$
pairs with $|m-n|=d$, excluding pairs for which $m$ or $n$ is closer
to the border than (typically) 6.  In our staggered formulation, even
and odd $d$ may correspond to different physical channels.

\begin{figure}[tbp]
\begin{center}
\leavevmode
\epsfig{file=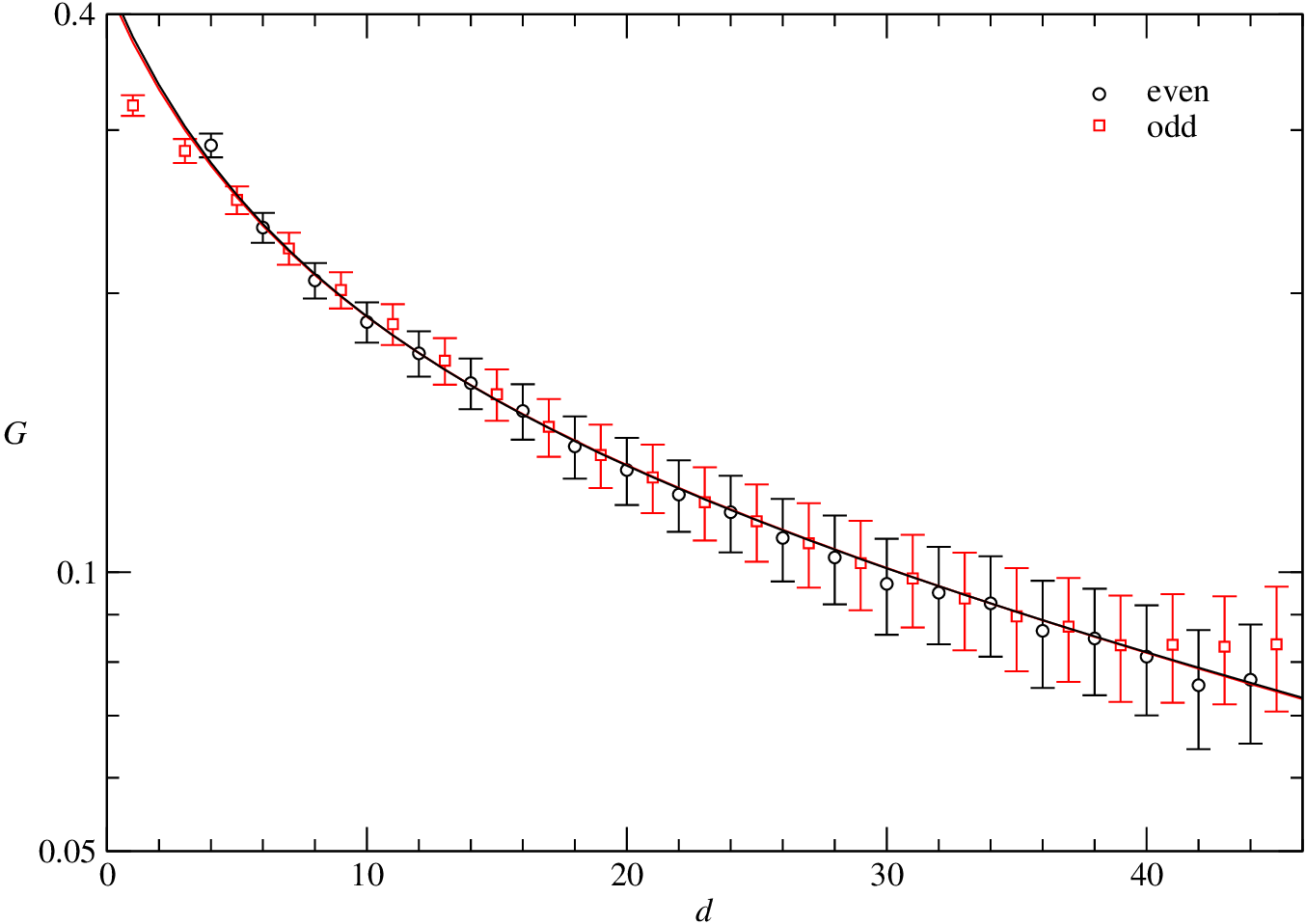,width=0.8\textwidth,angle=0}
\vskip-7mm
\end{center}
\caption{The connected correlation function $G_d$ for 
$V=0.5\,\varphi^2 - 0.5$.}
\label{fig:corr.l0-0.5}
\end{figure}

\begin{figure}[tbp]
\begin{center}
\leavevmode
\epsfig{file=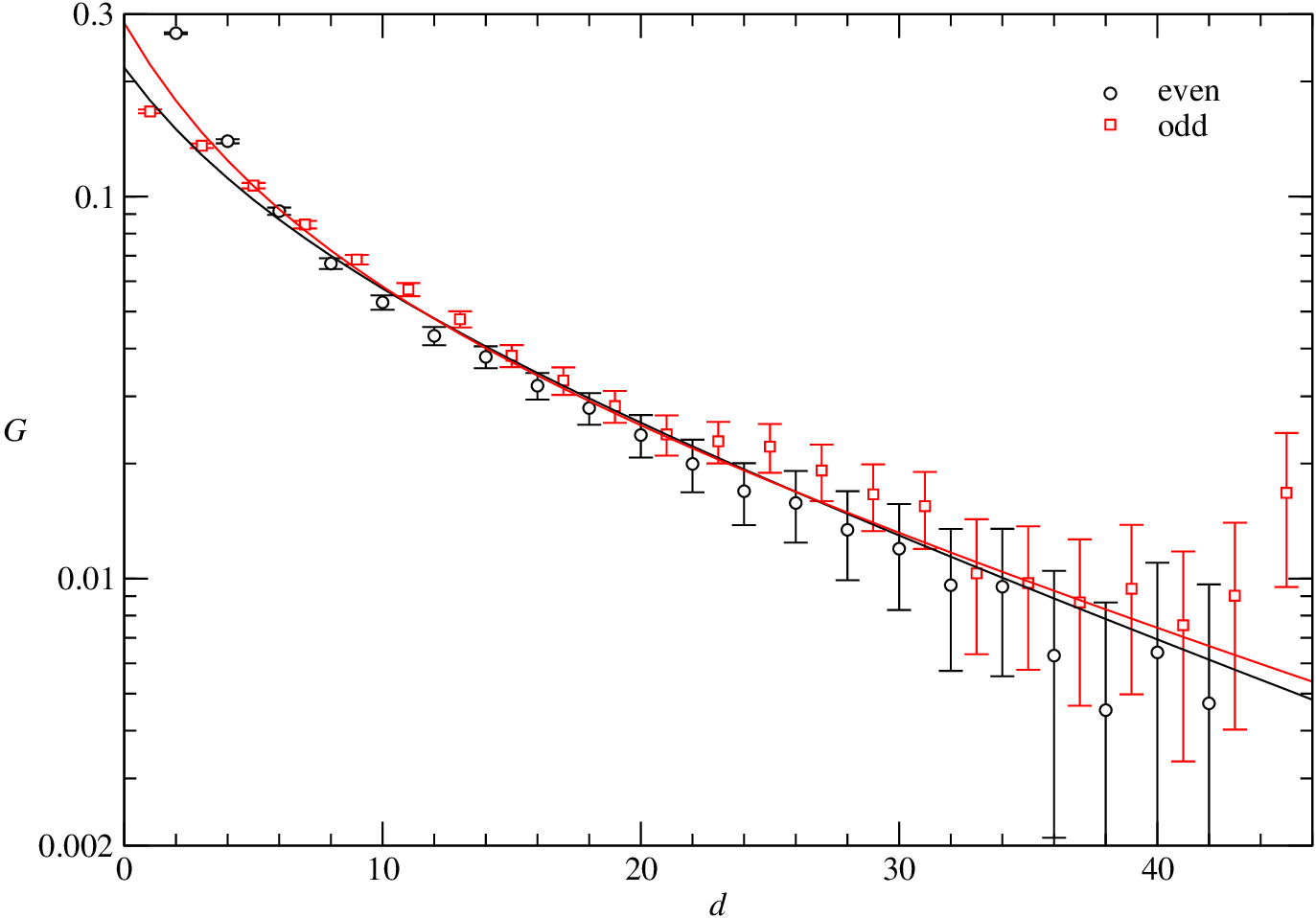,width=0.8\textwidth,angle=0}
\vskip-7mm
\end{center}
\caption{The connected correlation function $G_d$ for 
$V=0.5\,\varphi^2 - 0.38$.}
\label{fig:corr.l0-0.38}
\end{figure}

\begin{figure}[tbp]
\begin{center}
\leavevmode
\epsfig{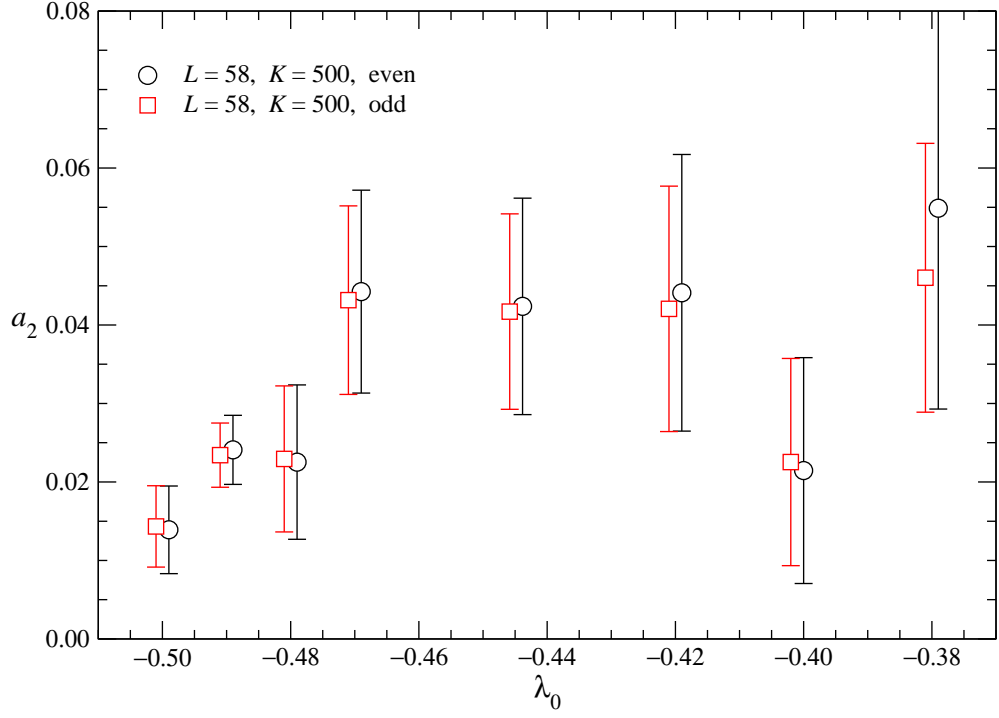}
\vskip-7mm
\end{center}
\caption{The effective mass $a_2$ of $G_d$ vs.\ $\lambda_0$ for
$L=58$ and $K=500$; for $\lambda_0\le-0.51$ the error on $a_2$ is very
large.}
\label{fig:a2}
\end{figure}

$G_d$ is fitted to the form $\exp[-a_1 - a_2 d + a_3/(d+10)]$,
separately for even and odd $d$; the best fits give a good $\chi^2$ if
we remove the smallest distances, typically $d\le3$ for the odd
channel and $d\le4$ for the even channel.  In the broken phase, we
have small but nonzero $a_2$, and we observe equivalence of the even-
and odd-$d$ channels; an example is shown in Fig.\ 
\ref{fig:corr.l0-0.5}.
In the unbroken phase, $a_2$ is larger, and the even- and odd-$d$
channels are somewhat different; an example is shown in Fig.\ 
\ref{fig:corr.l0-0.38}.
The difference between the two phases is apparent, e.g., in the plot
of $a_2$ vs.\ $\lambda_0$, presented in Fig.~\ref{fig:a2}; the
data presented here would indicate $\lambda_0^{(c)}=-0.48\pm0.01$.

An alternative window to the phase transition is offered by the
optimized values of the parameters of bosonic part of the trial wave
function, which should be related to the effective potential
$V_{\rm eff}$ of the bosonic field:
\[ V_{\rm eff}(\varphi) \sim - \alpha^B_4 \varphi^4 - \alpha^B_2 \varphi^2; \]
we verified that $\alpha^B_4<0$, as required by stability; a negative
value for $\alpha^B_2$ therefore indicates unbroken $Z_2$ symmetry,
while a positive $\alpha^B_2$ indicates spontaneous breaking of $Z_2$.

\begin{figure}[tbp]
\begin{center}
\leavevmode
\epsfig{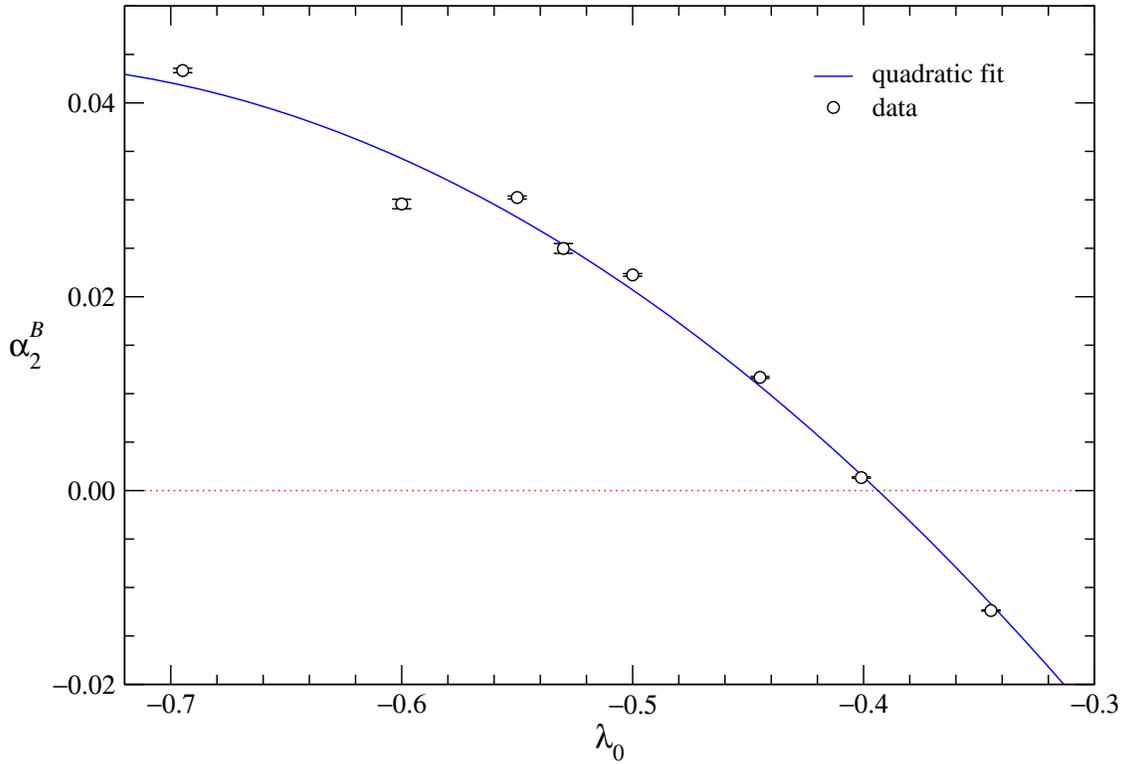}
\vskip-7mm
\end{center}
\caption{The optimization parameter $\alpha^B_2$; it is insensitive to
$L$ and $K$.}
\label{fig:alpha2B}
\end{figure}

The numerical values of $\alpha^B_2$ are shown in Fig.\ 
\ref{fig:alpha2B}.  It is clear that, especially in the broken phase,
statistical errors are underestimated.  The abovementioned scenario is
qualitatively confirmed, but the data yields
$\lambda_0^{(c)}\simeq-0.40$, which is very far from the more
traditional estimates obtained above.

\begin{figure}[tbp]
\begin{center}
\leavevmode
\epsfig{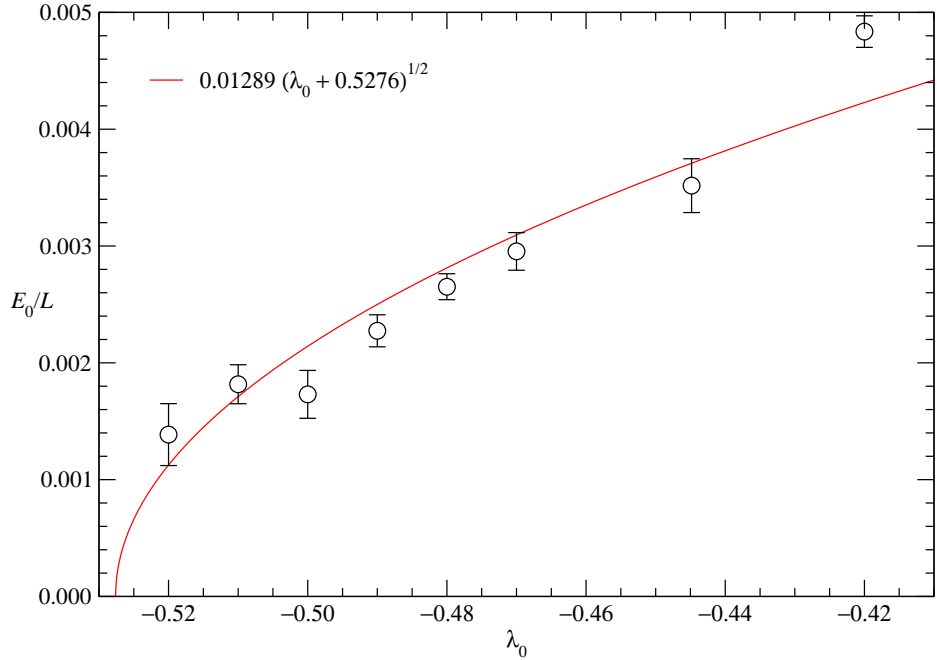}
\vskip-7mm
\end{center}
\caption{The ground-state energy density $E_0/L$ vs.\ $\lambda_0$.
The solid line is a fit to the form $E_0/L =
a\sqrt{\lambda_0-\lambda_0^{(c)}}$.}
\label{fig:E0vsl0}
\end{figure}

Finally, to investigate the supersymmetry properties of each phase, we
analyze ${\cal E}$, the ground-state energy density extrapolated to
infinite $K$ and $L$.  We fit $E_0/L$ to the form
\begin{equation}
{E_0 \over L} = {\cal E} + {a + b L \over K^\nu};
\label{eq:e0fit}
\end{equation}
$\chi^2/{\rm\#d.o.f}$ from 1 to 2; the errors on
the fit parameters are defined by the values giving an increase of
$\chi^2$ by 1; if $\chi^2>{\rm\#d.o.f}$ we multiply them by the
``scale factor'' $S = \sqrt{\chi^2/{\rm\#d.o.f}}$.  We present a plot
of ${\cal E}$ vs.\ $\lambda_0$ in Fig.~\ref{fig:E0vsl0}; these data
give $\lambda_0^{(c)}\sim-0.53$, with a rather large uncertainty.

\subsection{Continuum limit}
\label{Sec:continuum}

We wish to investigate if the pattern established in Sec.\
\ref{Sec:SSB} extends to the continuum limit.

We study the trajectory
\begin{equation}
\lambda_0={\lambda_2\over2\pi}\,\ln(4\lambda_2),
\label{eq:traj}
\end{equation}
corresponding to a 1-loop RG trajectory, cf.\ Eq.\
(\ref{eq:traj-latt}); the effect of $\lambda_0$ is
small in the range we considered, therefore we expect this to be
a reasonable approximation to a true RG trajectory.

We estimate the correlation length from the exponential decay of the
connected correlation function $G_d=\langle\varphi_n\varphi_{m}\rangle_c$
averaged over all $n,m$ pairs with $|m-n|=d$, excluding pairs for
which $m$ or $n$ is closer to the border than (typically) 8.
In our staggered formulation, even and odd $d$ correspond to
different physical channels.

\begin{figure}[tbp]
\begin{center}
\leavevmode
\epsfig{file=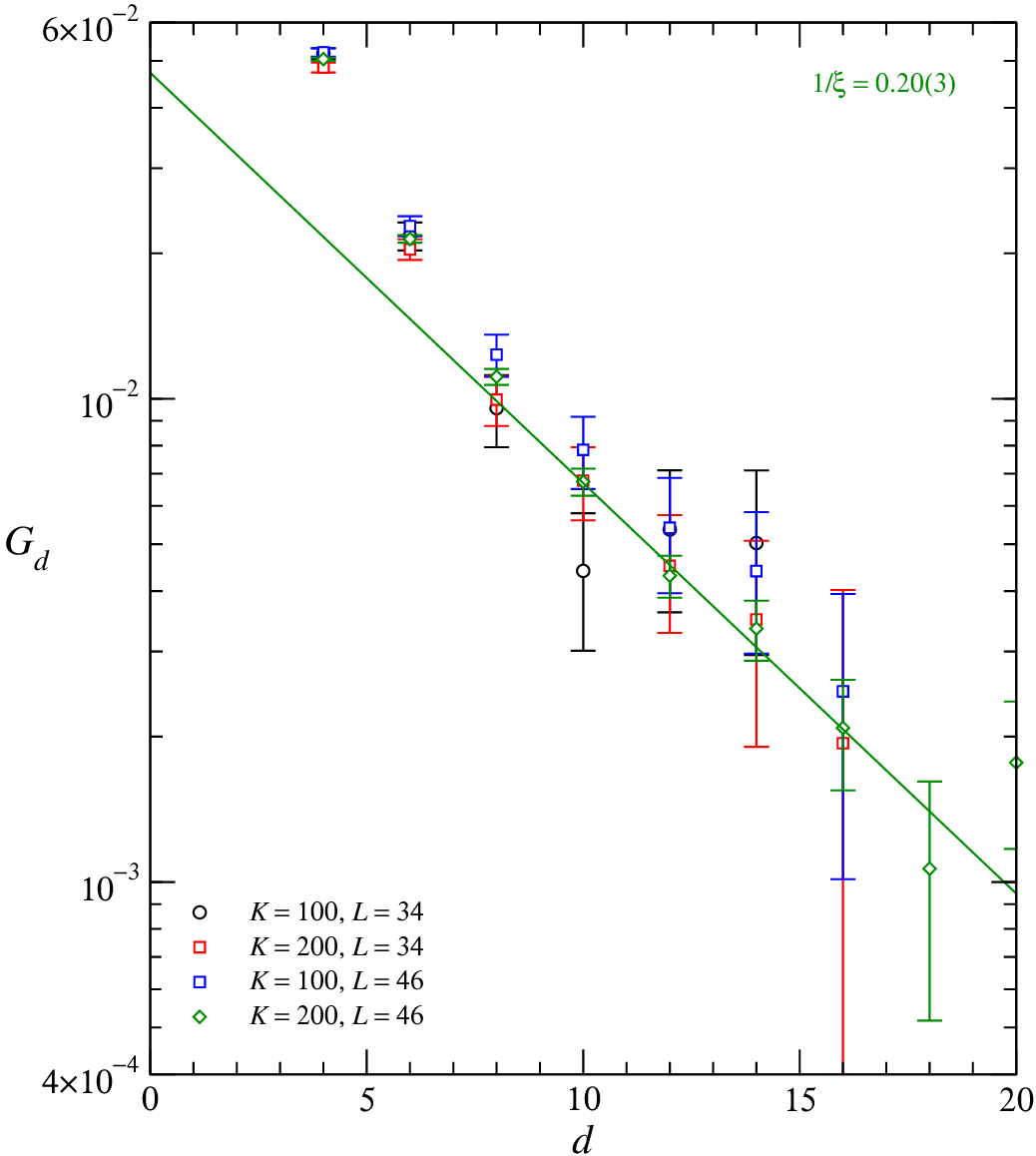,width=0.8\textwidth,angle=0}
\vskip-7mm
\end{center}
\caption{The connected correlation function $G_d$ at even distance for 
$V=0.353553\,\varphi^2 + 0.019502$; the curve and value of $1/\xi$
quoted are the result of an exponential fit for $10\le d\le18$ to the
$L=46$, $K=200$ data.}
\label{fig:corr.even}
\end{figure}

\begin{figure}[tbp]
\begin{center}
\leavevmode
\epsfig{file=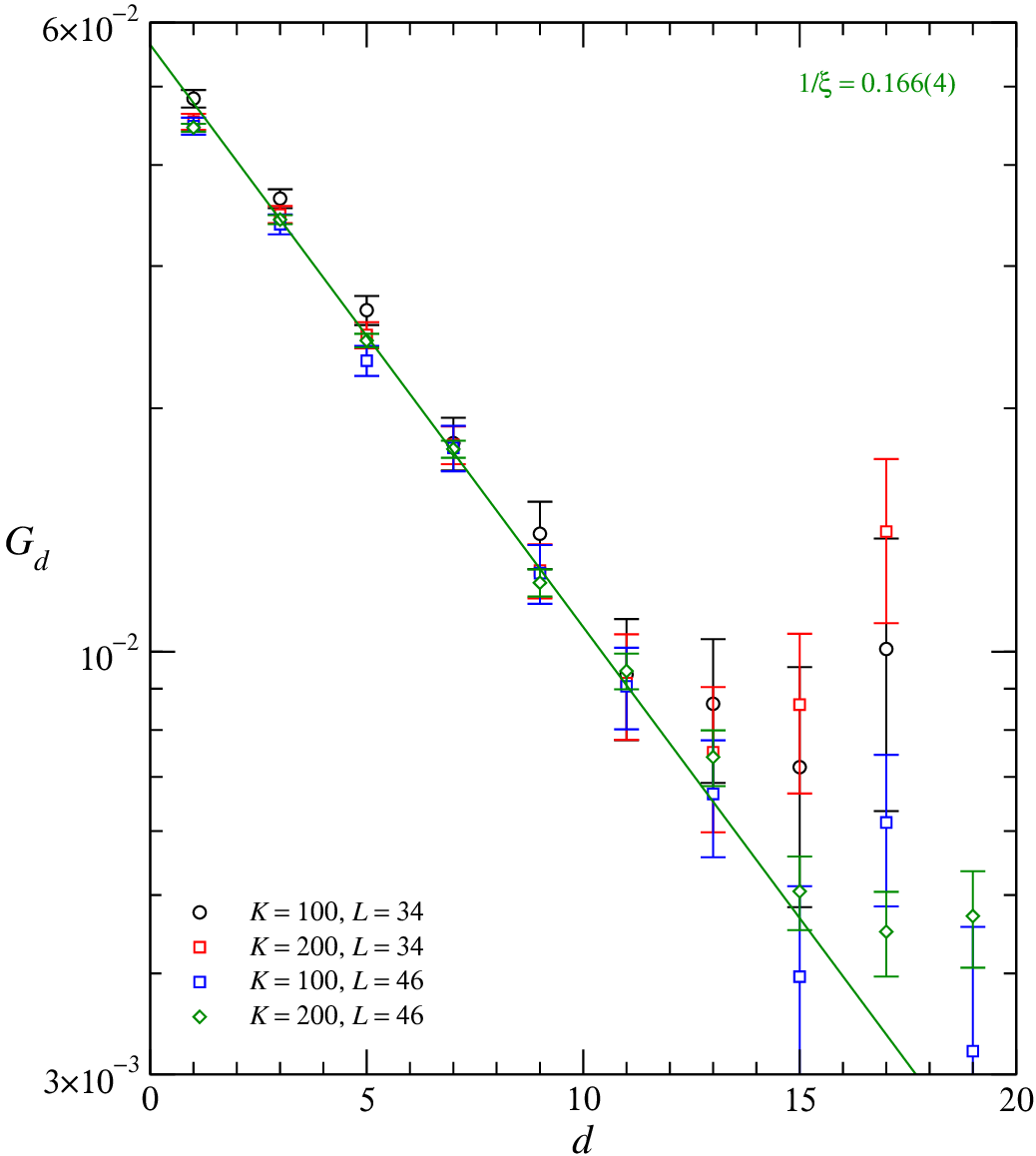,width=0.8\textwidth,angle=0}
\vskip-7mm
\end{center}
\caption{The connected correlation function $G_d$ at odd distance for 
$V=0.353553\,\varphi^2 + 0.019502$; the curve and value of $1/\xi$
quoted are the result of an exponential fit for $3\le d\le15$ to the
$L=46$, $K=200$ data.}
\label{fig:corr.odd}
\end{figure}

We performed runs for values of $\lambda_2$ spaced by a factor of
$\sqrt{2}$, with a statistics of $4{\times}10^6$ iterations.  In
Figs.\ \ref{fig:corr.even} and \ref{fig:corr.odd} we show the plots of
the $\varphi$ correlation for the case $V=0.35\,\varphi^2+0.02$.  It is very
difficult to extract a correlation length from the even-$d$ channel,
presumably because $\varphi$ has a very small overlap with the lightest
state of the channel, and the value $1/\xi=0.20\pm0.03$ quoted in
Fig.\ \ref{fig:corr.even} should be considered tentative.  The odd-$d$
channel is much cleaner, and it is possible to estimate $\xi$ with a
good precision.

\begin{figure}[tbp]
\begin{center}
\leavevmode
\epsfig{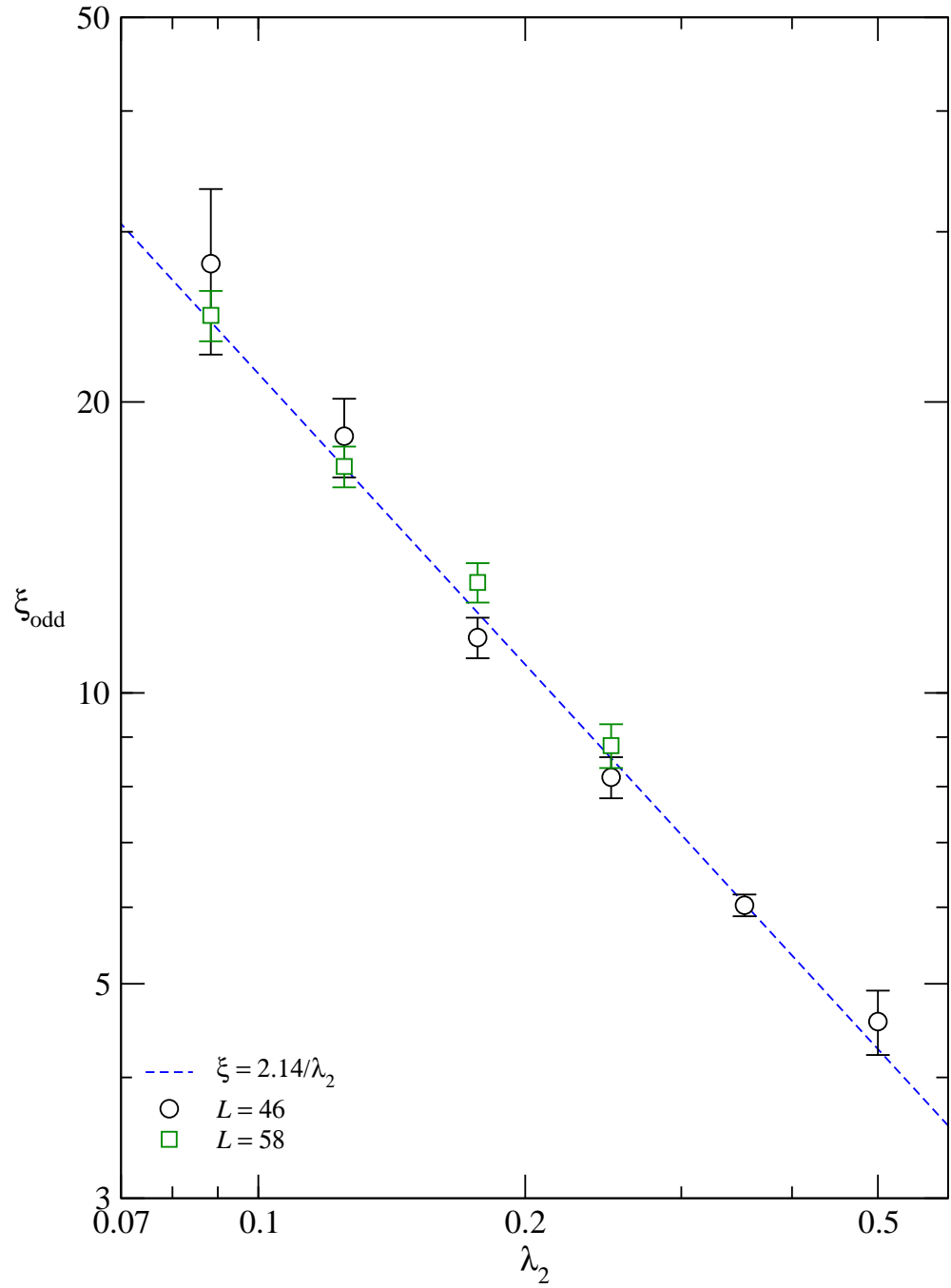}
\vskip-7mm
\end{center}
\caption{The correlation length at odd distance $\xi_{\rm odd}$.  The
dashed curve is the result of a scaling fit (with fixed exponent).}
\label{fig:xiodd}
\end{figure}

For the other values of $\lambda_2$, the situation is similar but with
slightly larger errors.  The measured values of $\xi_{\rm odd}$ follow
very well the na\"\i ve scaling behavior
\[
\xi\propto1/\lambda_2.
\]
The entire range $0.088\le\lambda_2\le0.35$ seems to be in the scaling
region, with $\lambda_2=0.5$ a borderline case, as shown in Fig.\ 
\ref{fig:xiodd}.  The values of $\xi_{\rm even}$ have very large
errors, and it is hard to draw any conclusion from them.

The Green Function Monte Carlo algorithm gives a very accurate
measurement of the ground-state energy $E_0$; to give a feeling of the
precision reached, we quote the results for the smallest and the
largest value of $\lambda_2$ we considered, along the trajectory
(\ref{eq:traj}):
\begin{eqnarray*}
E_0(\lambda_2{=}0.044,L{=}46,K{=}200)&=&(1.28\pm0.01){\times}10^{-3};\\
E_0(\lambda_2{=}0.5,L{=}46,K{=}200)&=&(69.44\pm0.05){\times}10^{-3}.
\end{eqnarray*}
$E_0$ show a sizable dependence on $K$ and $L$, and it is therefore
necessary to perform an extrapolation to $L\to\infty$ and
$K\to\infty$; we fitted the energy density to the form
\[
{E_0\over L} = {\cal E} + {c\over L} + {d\over L^2} + 
               K^\nu \left(e + {f\over L}\right),
\]
which gives a good $\chi^2$.  $\nu$ remains constant within errors,
with a value $\nu\approx0.75$, i.e., the algorithm performs well as we
approach the continuum limit.

\begin{figure}[tbp]
\begin{center}
\leavevmode
\epsfig{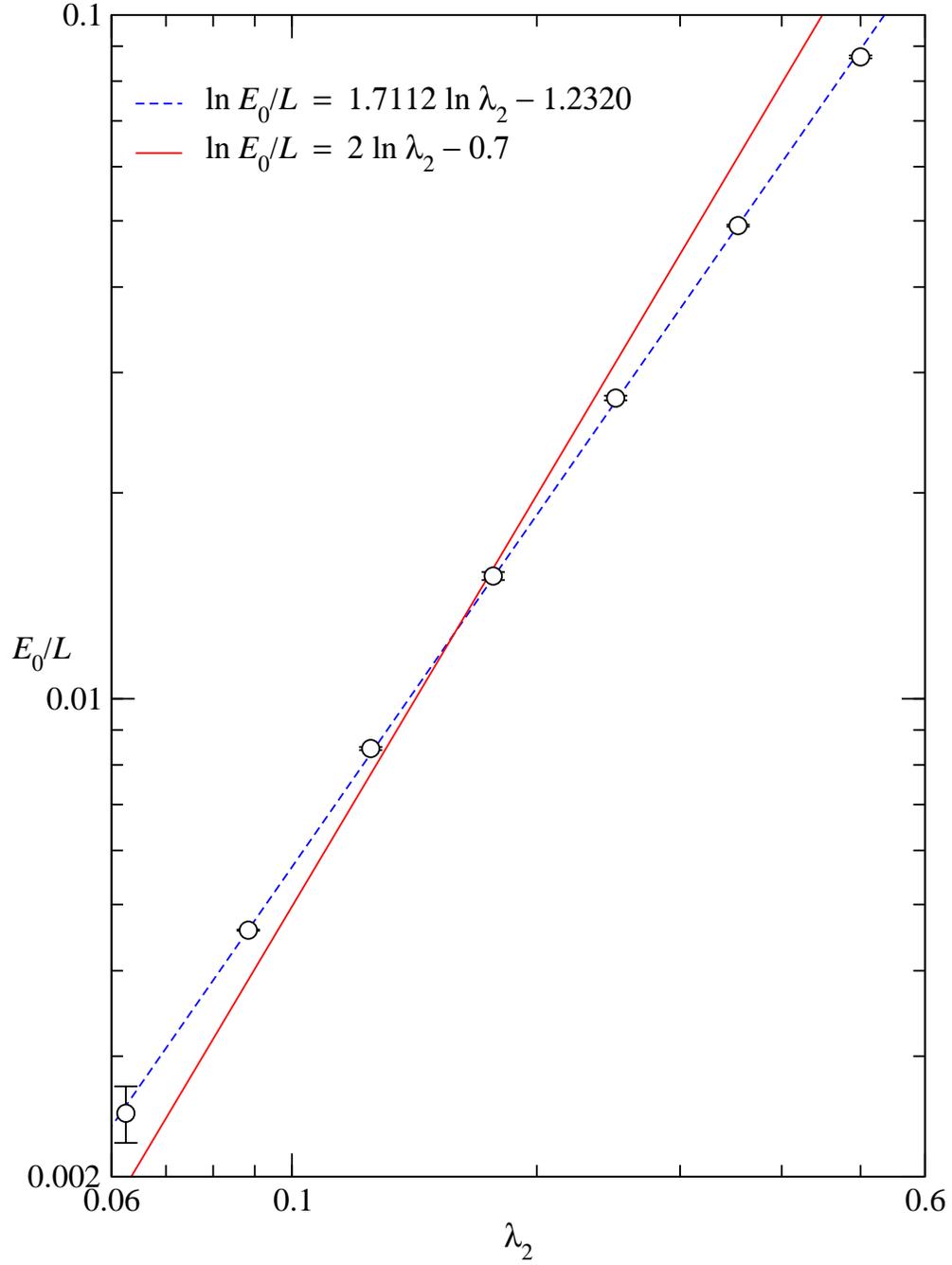}
\vskip-7mm
\end{center}
\caption{The ground-state energy density $E_0/L$, extrapolated to
$L\to\infty$ and $K\to\infty$.
The dashed curve is the result of a two-parameter fit, while the solid
curve shows the na\"\i ve scaling behavior.}
\label{fig:E0log}
\end{figure}

The ``scaling'' plot of the energy density $E_0/L$ is shown in Fig.\ 
\ref{fig:E0log}.  It seems to behave as $\lambda_2^{1.7}$, while
na\"\i ve scaling would predict $E_0/L\propto\lambda_2^2$.

The nonzero value of $E_0/L$ (disregarding this puzzling exponent) and
the lack of any signal for a breakdown of parity show that the
trajectory we are considering belongs to the phase with broken
supersymmetry and unbroken $Z_2$ symmetry.


\section{Conclusion}
\label{Sec:Conclusion}

In this paper, we investigated a class of two-dimensional $N=1$
Wess-Zumino models by non-perturbative lattice Hamiltonian techniques.
The key property of the formulation is the exact preservation of a
SUSY subalgebra at finite lattice spacing.  Our main tool are
numerical simulations using the Green Function Monte Carlo algorithm;
we also performed strong-coupling expansions.

All our results for the model with cubic prepotential indicate
unbroken supersymmetry.

We studied dynamical supersymmetry breaking in the model with
quadratic prepotential $V=\lambda_2\varphi^2 + \lambda_0$, performing
numerical simulations along a line of constant $\lambda_2$.  We
confirm the existence of two phases: a phase of broken SUSY and
unbroken $Z_2$ at high $\lambda_0$ and a phase of unbroken SUSY and
broken $Z_2$ at low $\lambda_0$, separated by a single phase
transition.

We studied the approach to the continuum limit in the model with
quadratic prepotential performing numerical simulations along a 1-loop
RG trajectory in the phase of broken supersymmetry; we measured the
correlation length of the bosonic field (in the odd-distance channel),
which is found to scale with the expected exponent; on the other hand,
the ground-state energy density scales with an exponent clearly
different from the expected exponent.

In many instances, the simulation algorithm suffers from slow
convergence in the number of walkers $K$.


\acknowledgments

It is a pleasure to thank Camillo Imbimbo, Ken Konishi, Gianni Morchio,
and Ettore Vicari for many helpful discussions and suggestions.


\appendix


\section{Check of unbroken SUSY for $\bm{V(\varphi) = \lambda_1\varphi + \lambda_0}$}
\label{App:LinearPotential}

If the potential $V(\varphi)$ is a linear function of the field $\varphi$,
then the ground state can be found explicitely. With a field translation
we can set $\lambda_0 = 0$. The model Hamiltonian is $H_B+H_F$ where we recall that
\begin{eqnarray}
H_B &=& \sum_{n=1}^L\left[\frac 1 2 p_n^2 + \frac 1 2 \left(\frac{\varphi_{n+1}-\varphi_{n-1}}{2}+\lambda_2
\varphi_n\right)^2\right] , \\
H_F &=& \sum_{n=1}^L\left[-\frac 1 2 (\chi_n^\dagger\chi_{n+1}+\chi_{n+1}^\dagger\chi_n) 
+(-1)^n \lambda_2 \chi_n^\dagger\chi_n\right] .
\end{eqnarray}
Thus, in the bosonic sector, the Hamiltonian can be written
\begin{equation}
H_B = \frac 1 2 \sum_n p_n^2 + \frac 1 2 \sum_{nm} \varphi_n V_{nm}^B \varphi_m ,
\end{equation}
with 
\begin{equation}
V_{nm}^B = \left\{\begin{array}{ll} 
\lambda_2^2+1/4 & n=m\ \mbox{and}\ n=1,L \\
\lambda_2^2+1/2 & n=m\ \mbox{and}\ 1<n<L \\
-1/4            & |n-m|=2
\end{array}\right.
\end{equation}
In the fermionic sector, the Hamiltonian can be written in terms of 
canonical Fermi annihilation and creation operators as 
\begin{equation}
H_F = \sum_{n,m} V_{nm}^F a_n^\dagger a_m
\end{equation}
with 
\begin{equation}
V_{nm}^F = \left\{\begin{array}{ll} 
(-1)^n\lambda_2 & n=m\\
\lambda_2^2+1/2 & n=m\ \mbox{and}\ 1<n<L \\
-1/2            & |n-m|=1
\end{array}\right.
\end{equation}
If we denote by $\{(\omega^B_n)^2\}$ and $\{\omega^F_n\}$ the {\em sorted\/} eigenvalues of 
$V^B$ and $V^F$, then we find that the ground state has actually zero energy
\begin{equation}
\label{cancellation}
E_0 = \frac 1 2 \sum_{n=1}^L\omega_n^B + \sum_{n=1}^{L/2}\omega^F_n = 0.
\end{equation}
This can be proved in the spirit of SUSY without computing explicitely the 
eigenvalues. In fact, we can check that $(V^F)^2$ is the matrix $V^B$ apart from a 
wrong sign in the diagonals $|n-m|=2$. This can be repaired by changing sign $\varphi\to -\varphi$
on the sites with $n\mathop{\rm mod}4 = 1,2$. Taking into account the particle-hole symmetry of $H_F$, 
we have thus proved that the spectra of $\sigma(V^F)$ and $\sigma(V^B)$ have the general form 
\begin{eqnarray}
\sigma(V^F) &=& \{-x_1, x_1, -x_2, x_2, \dots\} \\
\sigma(V^B) &=& \{x_1^2, x_1^2, x_2^2, x_2^2, \dots\} 
\end{eqnarray}
with full cancellation between the lowest $L/2$ fermionic values and one half of the 
square root of the bosonic ones as in Eq.~(\ref{cancellation}).


\section{Strong-coupling expansion of $\bm{\langle \varphi_k \rangle}$ and 
$\bm{\langle \varphi_k\varphi_l \rangle_c}$}
\label{App:First}

\subsection{$\bm{\langle \varphi_k \rangle}$}

Let us define 
\begin{equation}
\overline\varphi = \langle \varphi\rangle_+ = -\frac{\eta_0}{2\sqrt{2\varepsilon_0}}.
\end{equation}
The vacuum expectation value of the field $\varphi$ is 
\begin{equation}
\langle \gs | \varphi_k | \gs \rangle = \overline\varphi (-1)^k \langle \gs | (-1)^{n_k}| \gs \rangle = 
\overline\varphi (-1)^k (1-2\langle \gs | n_k| \gs \rangle) . 
\end{equation}
The expectation value of the occupation number can be computed by going to the basis $\{a\}$ and is
\begin{equation}
\langle \gs | n_k| \gs \rangle = \sum_{p=1}^{L/2} (v_k^{(p)})^2 .
\end{equation}
A straightforward calculation gives
\begin{equation}
\langle \gs | n_k| \gs \rangle = \frac{1}{2L}\left\{1+L+
\frac{\displaystyle\cos\left[\frac{\pi}{2L}(2k(L+1)-1\right]}{\displaystyle\sin\frac{\pi}{2L}(2k-1)}\right\},
\end{equation}
and therefore
\begin{equation}
\langle \gs | \varphi_k| \gs \rangle = \frac{\eta_0}{2\sqrt{2\varepsilon_0}}(-1)^k\frac{1}{L}\left\{1+
\frac{\displaystyle\cos\left[\frac{\pi}{2L}(2k(L+1)-1\right]}{\displaystyle\sin\frac{\pi}{2L}(2k-1)}\right\}.
\end{equation}
It is interesting to consider the limit $L\to\infty$ of this expression after a rescaling $k\to xL$ where 
$0<x<1$. The result is 
\begin{equation}
\langle \gs | \varphi_{xL}| \gs \rangle = \frac{\eta_0}{2\sqrt{2\varepsilon_0}}\left\{
\frac{1}{L}(\pm 1+\cot(\pi x)) + \frac{\pi}{2L^2}\frac{1}{\sin^2(\pi x)} + {\cal O}\left(\frac{1}{L^3}\right)\right\}
\end{equation}
where the sign is $+1$ for even $k=xL$ and $-1$ for odd $k$.

\subsection{$\bm{\langle \varphi_k\varphi_l \rangle_c}$}

Let us denote briefly
\begin{equation}
\langle A\rangle \equiv \langle\gs | A| \gs\rangle ,
\end{equation}
and
\begin{equation}
\langle AB\rangle_c = \langle AB\rangle - \langle A \rangle\langle B\rangle .
\end{equation}
The 2-point correlation is, for $k\neq l$,
\begin{equation}
\langle \varphi_k\varphi_l \rangle = (\overline\varphi)^2(-1)^{k+l}\langle  (-1)^{n_k+n_l} \rangle
= (\overline\varphi)^2(-1)^{k+l}(1- \langle n_k + n_k\rangle + 4\langle n_k n_l \rangle) .
\end{equation}
Going to the $\{a\}$ basis, we immediately obtain 
\begin{equation}
\langle n_k n_l\rangle_c = \sum_{1\le A \le L/2} v_k^A v_l^A \cdot \sum_{L/2+1\le B\le L} v_k^B  v_l^B ,
\end{equation}
and, for $k\neq l$,
\begin{equation}
\langle \varphi_k\varphi_l  \rangle_c = 4(\overline\varphi)^2(-1)^{k+l}\langle  n_k n_l \rangle_c .
\end{equation}
The two sums over eigenvalues can be evaluated analytically thanks to the simple form
of the eigenvectors $v_l^{(p)}$. The explicit result is 
\begin{eqnarray}
\frac{L}{2}\sum_{p=1}^{L/2} v_n^{(p)} v_m^{(p)} &=& \frac{L}{4}\delta_{n,m}+\frac{1}{4}Z_{n,m} , \\
\frac{L}{2}\sum_{p=1+L/2}^L v_n^{(p)} v_m^{(p)} &=& \frac{L}{4}\delta_{n,m}+
\frac{1}{2}(-1)^{n+m}-\frac{1}{4}Z_{n,m} ,
\end{eqnarray}
where
\begin{equation}
\normalbaselines
Z_{n,m} = \left\{ \begin{array}{lc}
n\ \mbox{even},\ m\ \mbox{even}: & \vphantom{\Biggl[}
(-1)^{\frac{n+m}{2}}\left[1+\cot\displaystyle\frac{\pi}{2L}(m+n-1)\right] \\
n\ \mbox{even},\ m\ \mbox{odd}: & \vphantom{\Biggl[}
(-1)^{\frac{n+m+1}{2}}\left[1+\cot\displaystyle\frac{\pi}{2L}(n-m)\right] \\
n\ \mbox{odd},\ m\ \mbox{even}: & \vphantom{\Biggl[}
(-1)^{\frac{n+m+1}{2}}\left[1+\cot\displaystyle\frac{\pi}{2L}(m-n)\right] \\
n\ \mbox{odd},\ m\ \mbox{odd}: & \vphantom{\Biggl[}
(-1)^{\frac{n+m}{2}}\left[-1+\cot\displaystyle\frac{\pi}{2L}(m+n-1)\right] 
\end{array}
\right. ,
\end{equation}
that can be used to compute the connected correlation on a finite lattice. 

It is interesting to note that 
the limit $L\to\infty$ can be taken without rescaling $n$ and $m$. For instance, we have
\begin{equation}
\lim_{L\to\infty} \langle \gs | \varphi_1\ \varphi_n | \gs\rangle_c = \frac{4(\overline\varphi)^2}
{\pi^2}(-1)^n \left\{\begin{array}{lc}
n\ \mbox{even}: & \vphantom{\Biggl[} \displaystyle \frac{1}{(n-1)^2} \\
n\ \mbox{odd}:  & \vphantom{\Biggl[} \displaystyle \frac{1}{n^2}
\end{array}\right. .
\end{equation}


\section{Second-order expansion for $\bm{E_0}$, even $\bm{q}$}
\label{App:Second}

The general formula for the second order contribution to the ground state energy is 
\begin{eqnarray}
E_2 &=& E_{2,1}+E_{2,2} \\
E_{2,1} &=&  \langle \Psi_0^{(1)} | H^{(4)} | \Psi_0^{(1)} \rangle \\
E_{2,2} &=& \sum_{\Psi'} \frac{
\langle \Psi_0^{(1)} | H^{(2)} | \Psi'\rangle \langle\Psi' | H^{(2)} | \Psi_0^{(1)} \rangle 
}{E_0-E'}
\end{eqnarray}
The states $|\Psi'\rangle$ are excited states of the form 
\[
|\Psi'\rangle = \psi_{k_1}^{\sigma_1}(\varphi_1)\cdots \psi_{k_L}^{\sigma_L}(\varphi_L)|n_1,\dots
n_L\rangle
\]
where $k_1+\cdots k_L = \nu > 0$ (integer) and $\sigma_l = (-1)^{n_l+l}$. For such a state we have
\[
E' = \sum_l \varepsilon_{k_l}
\]
A first important remark is that $H^{(2)}$ can be restricted to its $H^{(2)}_B$ part.
In fact, the fermionic part of $H^{(2)}$ can be written as $H^{(2)}_{eff}$ plus terms 
that are only functions of the occupation numbers. These operators 
acting on $|\Psi_0^{(1)}\rangle$ give states that are orthogonal to the subspace of 
excited unperturbed states $|\Psi'\rangle$.
\\
The first term in $E_2$ is simple and can be computed straightforwardly exploiting the fact that 
the expectation value of $\varphi^2_l$ over $|\Psi^{(1)}\rangle$ does not depend on $l$:
\begin{equation}
E_{2,1} = \frac{L-1}{4}\overline{\varphi^2} -\frac{1}{4}\sum_{l=1}^{L-2}\langle \Psi_0^{(1)} |
\varphi_l\varphi_{l+2} | \Psi_0^{(1)}\rangle 
= \frac{L-1}{4}\overline{\varphi^2}  -\frac{1}{4} (\overline\varphi)^2 c_0(L)
\end{equation}
where 
\[
\overline{\varphi^2} = \int_{-\infty}^\infty \varphi^2 [\psi_0^\pm(\varphi)]^2 d\varphi
\]
and 
\begin{equation}
c_0(L) =  \sum_{l=1}^{L-2} (1- \langle n_l\rangle -\langle n_{l+2}\rangle + 
4\langle n_l n_{l+2}\rangle ) 
\end{equation}
\\
About $E_{2,2}$, it can be computed by considering states with an excited single-site wave function
in one or two distinct sites. Summing the two contributions we find 
\begin{eqnarray}
E_2 &=& \frac{L-1}{4}\overline{\varphi^2}-\frac{1}{4}
(\overline\varphi)^2 c_0(L) \nonumber \\
&-& \sum_{s>0}\frac{1}{\varepsilon_s-\varepsilon_0}\left(\frac{1}{2}\left(\Phi^{(1)}_s\right)^2 \left(V^{(0)}\right)^2 +c_1(L)\ \Phi^{(1)}_s\ V^{(1)}_s\ \overline\varphi\ 
V^{(0)} + c_2(L)\ 
(\overline\varphi)^2 \left(V^{(1)}_s\right)^2 \right) \nonumber \\
&-& \sum_{s>0,\,t>0}\frac{1}{\varepsilon_s+\varepsilon_t-2\varepsilon_0}\Biggl\{
\frac{1}{4}(L-1)
\left[\left(\Phi^{(1)}_s\right)^2 \left(V^{(1)}_t\right)^2
+\left(\Phi^{(1)}_t\right)^2 \left(V^{(1)}_s\right)^2\right] \nonumber \\
&&\qquad\qquad\qquad\qquad\quad+\,
c_3(L) \Phi^{(1)}_s \Phi^{(1)}_t V^{(1)}_s V^{(1)}_t
\Biggr\}
\end{eqnarray}
The symbols appearing in the above equations are defined as follows
\begin{eqnarray}
V^{(0)} &=& \langle\psi_0^\pm | V(\varphi) | \psi_0^\pm \rangle = \sqrt{2\varepsilon_0}\eta_0 \\
V^{(1)}_s    &=& \frac{1}{\sqrt{2}}(\sqrt{\varepsilon_0}+(-1)^s \sqrt{\varepsilon_s})
                    \langle \psi_0^- | \psi_s^+\rangle, \\
\Phi^{(1)}_s &=& -\frac{1}{\sqrt{2}}\frac{1}{\sqrt{\varepsilon_0}+(-1)^s \sqrt{\varepsilon_s}}
                             \langle \psi_0^- | \psi_s^+\rangle
\label{defend}
\end{eqnarray}
The functions $c_{0,1,2,3}(L)$ can be fitted with a few 
powers of $L$ and the numerical result is
\begin{eqnarray}
c_0(L) &=& L-1.495-\frac{2.610}{L}+\cdots \\
c_1(L) &=& -0.540-\frac{1.698}{L}+\frac{1.787}{L^2}+\cdots  \\
c_2(L) &=& \frac 1 2 L -0.540+\frac{0.301}{L}+\frac{1.788}{L^2}+\cdots \\
c_3(L) &=& -\frac{2}{\pi^2} L -\frac{0.167}{L} +\cdots
\end{eqnarray}
The full expression for $E_2$ has thus the following simpler asymptotic expression 
\begin{eqnarray}
\lim_{L\to\infty}\frac{E_2}{L} &=& \frac{1}{4} (\overline{\varphi^2}-\overline\varphi^2)  
 -\frac{1}{2}(\overline\varphi)^2 \sum_{s>0}\frac{\left(V^{(1)}_s\right)^2}{\varepsilon_s-\varepsilon_0}
 + \nonumber \\
&-& \sum_{s>0, t>0}\frac{1}{\varepsilon_s+\varepsilon_t-2\varepsilon_0}\left\{
\frac{1}{4}
\left[\left(\Phi^{(1)}_s\right)^2 \left(V^{(1)}_t\right)^2
+\left(\Phi^{(1)}_t\right)^2 \left(V^{(1)}_s\right)^2\right] + \right. \nonumber\\
&&\left.\qquad-\, \frac{2}{\pi^2} 
\Phi^{(1)}_s \Phi^{(1)}_t V^{(1)}_s V^{(1)}_t
\right\} .
\end{eqnarray}
As an example, we find for the model with $V(\varphi)=\varphi^2$:
\[
E(\lambda^2, 22) = 0.2811-\frac{0.160}{\lambda^2}+\frac{0.0481}{\lambda^4},
\ \mbox{and}\ 
E(\lambda^2, \infty) = 0.2811-\frac{0.160}{\lambda^2}+\frac{0.0488}{\lambda^4}
\]
If we are interested in a comparison with an actual simulation, some trivial 
rescaling is necessary. For instance, in the case of a purely quadratic 
$V(\varphi) = \lambda_2\ \varphi^2$ we must compare the expansion 
and the results for $\lambda_2^{-2/3} E_0/L$ and identify $\lambda = \lambda_2^{1/3}$ as
the strong-coupling expansion parameter.


\section{Factorized wave function for the free model}
\label{App:Free}

In the free case $V\equiv 0$, with even $L$, we have 
\begin{equation}
H^{(0)} = H_B + H_F, 
\end{equation}
where ($\varphi_0 = \varphi_{L+1} \equiv 0$, the same for fermions)
\begin{eqnarray}
H_B &=& \sum_{n=1}^L \left\{\frac 1 2 p_n^2+\frac 1 8
(\varphi_{n+1}-\varphi_{n-1})^2 \right\}, \\
H_F &=& -\frac 1 2 \sum_{n=1}^L (\chi_n^\dagger \chi_{n+1} + 
\chi^\dagger_{n+1}\chi_n) .
\end{eqnarray}

The two Hamiltonians $H_B$ and $H_F$ are decoupled and the ground
state takes the factorized form 
\begin{equation}
|\Psi^{(0)}\rangle = |\Psi^{(0)}_B\rangle \otimes |\Psi^{(0)}_F\rangle .
\end{equation}
We now determine $|\Psi^{(0)}_{B, F}\rangle$ separately, assuming 
$L\mathop{\rm mod}4 = 2$ in order to have a unique ground state
in the decoupled $V\equiv 0$ model.

\subsection{Bosonic sector}

Let us write $H_B$ in the form 
\begin{equation}
H_B = \frac 1 2 \sum_{n=1}^L p_n^2 + \frac 1 2 \sum_{n,m=1}^L
V_{nm}\varphi_n\varphi_m .
\end{equation}
Let $\lambda^2_k$ be the eigenvalues of the $L\times L$ matrix
$V$ and let $z^{(k)}_n$ be 
the corresponding (real) eigenvectors satisfying 
\begin{equation}
z^{(k)}\cdot z^{(l)} = \delta_{k,l} .
\end{equation}
The ground state is
\begin{equation}
\langle \varphi | \Psi^{(0)}_B\rangle = 
\exp\left(-\frac 1 2 \sum_{n,m=1}^L R_{nm}\varphi_n\varphi_m\right),
\end{equation}
where
\begin{equation}
R_{nm} = \sum_{k=1}^L \lambda_k z_n^{(k)} z_m^{(k)}
\end{equation}
The explicit form of $\lambda_k$ is
\begin{equation}
\lambda_k = \sin\left(\frac{2i-1}{L+1}\frac\pi 2\right), \quad
i=\lfloor\half(k+1)\rfloor \quad (i = 1, \dots, \half L);
\end{equation}
the dimension of each eigenspace is 2.
The ground state energy of $H_B$ is
\begin{equation}
E_{0,B} = \sum_{k=1}^L \half\lambda_k = 
\sum_{i=1}^{L/2} \sin\left(\frac{2i-1}{L+1}\frac\pi 2\right).
\end{equation}

We adopt for the two orthogonal eigenvectors the choice
\begin{equation}
z^{(2i-1)}_n = \frac{2}{\sqrt{L+1}} \left\{\begin{array}{ll}
0 & \mbox{($n$ even)} \\
\displaystyle
\sin\left(\frac{2i-1}{L+1}\frac\pi 2 (L-n+1)\right) & \mbox{($n$ odd)}
\end{array}\right.
\end{equation}
\begin{equation}
z^{(2i)}_n = \frac{2}{\sqrt{L+1}} \left\{\begin{array}{ll}
\displaystyle
\sin\left(\frac{2i-1}{L+1}\frac\pi 2 n\right) & 
\mbox{($n$ even)} \\
0 & \mbox{($n$ odd)} \\
\end{array}\right.
\end{equation}

\subsection{Fermionic sector}

In the free case, the Hamiltonian in the fermionic sector is
diagonalized by the orthogonal change of basis
\begin{equation}
\chi_x = \sqrt\frac{2}{L+1}\sum_{n=1}^{L}\sin\frac{n\pi x}{L+1} a_n,
\end{equation}
and takes the form 
\begin{equation}
\label{xxx}
H_F = -\sum_{n=1}^{L}\cos\frac{n\pi}{L+1} a_n^\dagger a_n,
\end{equation}
Hence the one-fermion energies are 
\begin{equation}
-\cos\frac{n\pi}{L+1},\quad  n=1,\dots, L .
\end{equation}
The fermionic component of the (supersymmetric) ground state is
\begin{equation}
|\Psi^{(0)}_F\rangle  = \prod_{n=1}^{L/2} a_n^\dagger |0\rangle ;
\end{equation}
the ground state energy of $H_F$ is
\begin{equation}
E_{0,F} = -\sum_{i=1}^{L/2} \cos\frac{n\pi}{L+1};
\end{equation}
we can easily check that 
\begin{equation}
E_{0,F} = -E_{0,B} = \frac 1 2 \left(1-\frac{1}{\displaystyle \sin \frac{\pi}{2(L+1)}}\right),
\end{equation}
and therefore $E_0 = E_{0,B} + E_{0,F} = 0$.


\end{document}